\documentclass[a4paper,11pt]{article}
\pdfoutput=1

\usepackage{Class/jcap}
\usepackage{amsmath,amssymb,bm,float,Misc/Macro,mathrsfs,tabularx,url}
\usepackage{threeparttable}
\usepackage{xcolor}
\usepackage{fontawesome5}

\def\hpol{\widehat{p}}

\def\pteppaper{LB23}
\def\paperII{LB-Delensing}

\def\LB{{\it LiteBIRD}}
\def\Planck{{\it Planck}}
\def\Euclid{{\it Euclid}}

\author[1,5,6]{A.\,I.\,Lonappan,}
\author[2]{T.\,Namikawa,}
\author[1]{G.\,Piccirilli,}
\author[3,4]{P.\,Diego-Palazuelos,}
\author[3,4]{M.\,Ruiz-Granda,}
\author[1,5]{M.\,Migliaccio,}
\author[6,7,8]{C.\,Baccigalupi,}
\author[9,10,11]{N.\,Bartolo,}
\author[12]{D.\,Beck,}
\author[13]{K.\,Benabed,}
% ...
\affiliation[1]{Dipartimento di Fisica, Università di Roma Tor Vergata, Via della Ricerca Scientifica, 1, 00133, Roma, Italy}
\affiliation[2]{Kavli Institute for the Physics and Mathematics of the Universe (Kavli IPMU, WPI), UTIAS, The University of Tokyo, Kashiwa, Chiba 277-8583, Japan}
\affiliation[3]{Instituto de Fisica de Cantabria (IFCA, CSIC-UC), Avenida los Castros SN, 39005, Santander, Spain}
\affiliation[4]{Dpto. de Física Moderna, Universidad de Cantabria, Avda. los Castros s/n, E-39005 Santander, Spain}
\affiliation[5]{INFN Sezione di Roma2, Università di Roma Tor Vergata, via della Ricerca Scientifica, 1, 00133 Roma, Italy}
\affiliation[6]{International School for Advanced Studies (SISSA), Via Bonomea 265, 34136, Trieste, Italy}
\affiliation[7]{INFN Sezione di Trieste, via Valerio 2, 34127 Trieste, Italy}
\affiliation[8]{IFPU, Via Beirut, 2, 34151 Grignano, Trieste, Italy}
\affiliation[9]{Dipartimento di Fisica e Astronomia “G. Galilei”, Universita` degli Studi di Padova, via Marzolo 8, I-35131 Padova, Italy}
\affiliation[10]{INFN Sezione di Padova, via Marzolo 8, I-35131, Padova, Italy}
\affiliation[11]{INAF, Osservatorio Astronomico di Padova, Vicolo dell’Osservatorio 5, I-35122, Padova, Italy}
\affiliation[12]{Stanford University, Department of Physics,  CA 94305-4060, USA}
\affiliation[13]{Institut d'Astrophysique de Paris, CNRS/Sorbonne Université, Paris France}
% ...

\emailAdd{anto.lonappan@roma2.infn.it}

\abstract{
We explore the capability of measuring lensing signals in \LB\ full-sky polarization maps. 
With a $30$ arcmin beam width and an impressively low polarization noise of $2.16\,\mu$K-arcmin, \LB\ will be able to measure the full-sky polarization of the cosmic microwave background (CMB) very precisely. This unique sensitivity also enables the reconstruction of a nearly full-sky lensing map using only polarization data, even considering its limited capability to capture small-scale CMB anisotropies. In this paper, we investigate the ability to construct a full-sky lensing measurement in the presence of Galactic foregrounds, finding that several possible biases from Galactic foregrounds should be negligible after component separation by harmonic-space internal linear combination. We find that the signal-to-noise ratio of the lensing is approximately $40$ using only polarization data measured over $90\%$ of the sky. This achievement is comparable to \Planck's recent lensing measurement with both temperature and polarization and represents a four-fold improvement over \Planck's polarization-only lensing measurement.
The \LB\ lensing map will complement the \Planck\ lensing map and provide several opportunities for cross-correlation science, especially in the northern hemisphere. 
} 

\title{LiteBIRD Science Goals and Forecasts: A full-sky measurement of gravitational lensing of the CMB}

\begin{document}
\maketitle
\flushbottom

\section{Introduction}

Cosmic microwave background (CMB) polarization has been measured by multiple CMB experiments to improve constraints on cosmology. A CMB linear polarization map contains two spatial patterns: parity-even $E$-modes and parity-odd $B$-modes  \cite{Zaldarriaga:1996:EBdef,Kamionkowski_1997}. 
In linear theory, density perturbations are the dominant source of the temperature anisotropies and $E$-mode polarization. The density perturbations do not produce the $B$-mode polarization without non-linear effects \cite{Seljak:1996ti}. However, inflationary gravitational waves could generate the $B$-mode polarization \cite{Seljak:1996gy,Kamionkowski:1996zd}. 
The main goal of the \LB\ experiment is the measurement of $B$-mode polarization produced by inflationary gravitational waves, which would be considered conclusive evidence for inflation in the early Universe \cite{LiteBIRD}.

Another effect that induces $B$-mode polarization is the weak gravitational lensing of the CMB. The mass distribution in the late Universe disturbs the trajectory of CMB photons, which distorts the spatial pattern of the observed polarization maps and converts part of the $E$-mode polarization into $B$-mode polarization \cite{Zaldarriaga:1998:LensB}. The gravitational lensing distortion is a nonlinear effect on the CMB.
A measurement of CMB lensing allows us to learn about the matter distribution in the late Universe. The lensing mass distribution correlates with tracers of the large-scale structure. Such correlations have been measured using, e.g., the galaxy number density \cite{Smith:2007,Hirata:2008,Bianchini:2014,Giannantonio:DES:2015,Singh:2016xey,Omori:DES:2018:gal,Polarbear:2019:gal,Marques:2019,Darwish:2020,Krolewski:2021,Dong:2021:void,Sun:2021,Miyatake:2021,Saraf:2021,Omori:DES:2022,Chang:DES:2022,Piccirilli:2022:phixgal,Yao:2023,Farren:ACT:2023,Farren:2023}, cosmic shear \cite{Hand:2013:phixshear,Kirk:DES:2015,Liu:2015,Harnois-Deraps:2016,Singh:2016xey,Harnois-Deraps:2017kfd,Omori:DES:2018:shear,Namikawa:2019:hscxpb,ACT:Robertson:2020,Marques:2020,Chang:DES:2022,Shaikh:ACT:2023}, the late-time integrated Sachs-Wolfe (ISW) effect in the CMB temperature fluctuations \cite{P13:ISW,P15:ISW,P18:phi,Carron:2022:NPIPE-phixISW}, the thermal Sunyaev-Zel'dovich effect \cite{Hill:2013,McCarthy:2023:tSZ}, and the cosmic infrared background \cite{Holder:2013:phixCIB,P13:CIBxphi,Hanson:2013:lensB,PB:2014:phixI,vanEngelen:ACT:2014:cib,P18:phi,Cao:2019:phixI}. 
These cross-correlations have been used to constrain cosmology. 
In addition, the lensing map helps to improve the statistical uncertainty of the inflationary gravitational waves with so-called ``delensing'' \cite{Kesden:2002:delens,Seljak:2003pn,Smith:2010:delens}. 

The lensed CMB polarization data has off-diagonal correlations between angular multipoles, which can be utilized to reconstruct the gravitational lensing potential \cite{OkamotoHu:quad,HuOkamoto:2001}. 
Multiple CMB experiments have reconstructed the CMB lensing mass map. Its angular power spectrum has been measured, by the Atacama Cosmology Telescope \cite{Das:2011,Sherwin:2011:DE,ACT16:phi,Qu:ACT:2023,Madhavacheril:ACT:2023,MacCrann:ACT:2023}, BICEP \cite{BKVIII,BICEPKeck:2022:los-dist}, \Planck\ \cite{P13:phi,P15:phi,P18:phi,Carron:2022:NPIPE-lensing}, POLARBEAR \cite{PB:phi:2013,PB:phi:2019}, and the South Pole Telescope \cite{SPT:phi:2012,Story:2015,SPT:phi:2019,Millea:2020,Pan:SPT:2023}. Upcoming and future ground-based CMB experiments, including the Simons Observatory \cite{SimonsObservatory}, CMB-S4 \cite{CMBS4} and Ali CMB Polarization Telescope (AliCPT) \cite{AliCPT:phi}, are planning to make high-sensitivity measurements of CMB polarization. Their target sensitivities are enough to measure the small-scale $B$-mode signal caused by lensing and significantly improve the precision of lensing measurements. After \Planck\, however, only \LB\ can reconstruct a full-sky lensing mass map including both the northern and southern hemispheres, which is impossible to achieve with a single ground-based experiment. Therefore, the \LB\ lensing mass map provides an exciting opportunity to cross-correlate the CMB lensing with galaxies on the sky. The \LB\ lensing map in the northern hemisphere would also be complementary with the AliCPT lensing map \cite{AliCPT:phi}. 

This work is part of a series of papers that present the science achievable by the \LB\ space mission, expanding on the overview of the mission published in Ref.~\cite{LiteBIRD} (hereafter, \pteppaper). 
In particular, this work focuses on the initial investigation of the capability of measuring lensing signals with \LB. 
Multiple studies have explored the practical issues and developed mitigation techniques for these issues, such as survey boundary and source masking \cite{Perotto:2009,Plaszczynski:2012,Namikawa:2012:bhe,namikawa-MF,BenoitLevy:2013:cutsky}, inhomogeneous noise \cite{Namikawa:2012:bhe,Hanson:2009:noise}, extragalactic foregrounds \cite{namikawa-MF,Osborne:2013nna,Schaan:2018,Mishra:2019,Sailer:2020,Han:2021:FG,Darwish:2021,Sailer:2022,Qu:2022:shear}, and instrumental systematics on lensing measurements \cite{Mirmelstein:2020:sys,Nagata:2021:sys}. Among practical concerns on lensing analysis, the Galactic foregrounds are one of the most important issues for \LB. Reference \cite{Beck:2020:lens-FG} shows that the residual Galactic foregrounds should be negligible in CMB-S4-like experiments that utilize small-scale CMB anisotropies. However, we use large-scale CMB polarization data in the \LB\ lensing measurement, where the Galactic foregrounds could be impactful \cite{Beck:2020:lens-FG}. 
This paper addresses this question by simulating the lensing reconstruction from a nearly full-sky \LB\ polarization map, including the component-separation procedure. We further investigate some applications of the \LB\ lensing map, including the primordial non-Gaussianity, the late-time ISW effect, and inflationary gravitational waves. 

This paper is organized as follows. In Section ~\ref{sec:reconst}, we explain our method for lensing reconstruction for \LB. In Sec.~\ref{sec:simulation}, we describe our setup for simulation. In Sec.~\ref{sec:results}, we show our main results for the lensing reconstruction for \LB. In Sec.~\ref{sec:applications}, we investigate potential applications of the \LB\ lensing map for cosmology. Finally, Sec.~\ref{sec:summary} is devoted to a summary and discussion. Throughout this paper, we use the cosmological parameters for the flat $\Lambda$CDM model adopted in \pteppaper. In a companion paper \cite{LiteBIRD:lens:paper2} (hereafter \paperII), we explore the feasibility of delensing for \LB. We choose the cosmological parameters obtained from \cite{P18:cosmological-parameters}. 

\section{Methodology} \label{sec:reconst}

This section describes our methodology for the lensing reconstruction from \LB\ polarization data. 
After briefly reviewing the principal impacts of weak gravitational lensing on CMB anisotropies, we summarize the quadratic estimator for the lensing reconstruction \cite{HuOkamoto:2001,OkamotoHu:quad}, the internal-linear combination (ILC) in harmonic space (hereafter HILC) \cite{Tegmark:2003} for component separation, and filtering of the polarization map. 
The temperature map obtained from \LB\ will not offer significant improvement over the results already obtained by the \Planck\ mission due to its restricted angular resolution. However, \LB\ will be able to precisely measure CMB polarization over the entire sky, providing a complementary full-sky lensing map that will have better accuracy at large angular scales ($L<10$) since the $EB$ quadratic estimator will be much less sensitive to several potential mean fields \cite{namikawa-MF}. Our study focuses on the polarization measurements of \LB, where the polarization quadratic estimators yield the highest signal-to-noise ratio (SNR). The rest of this article will thus concentrate on polarization analysis. 

\subsection{Gravitational Lensing of CMB}

The trajectories of CMB photons passing through a mass distribution are deflected by the gravitational potential which is referred to as gravitational lensing. The gravitational potential of the large-scale structure in the late-time Universe causes the gravitational lensing effect on the CMB, and the lensing effect distorts the observed CMB anisotropies. The lensed polarization anisotropies described by the Stokes $Q$ and $U$ parameters that we measure today are approximated as (e.g., Refs.~\cite{Challinor:2002cd,Lewis:2006:review,Hanson:2009:review})
\al{
    \left[\Tilde{Q} \pm i \Tilde{U}\right] (\hatn) &= \left[Q \pm i U\right] (\hatn + \bm{d})
    \,, \label{eq:tp_deflect}
}
where $\hatn$ is the line-of-sight unit vector and $\bm{d}$ is the deflection vector. We define the lensing potential, $\phi$, as $\bm{d}=\bm{\nabla}\phi$, where $\bm{\nabla}$ is the covariant derivative on the sphere. The lensing potential is related to the line-of-sight projection of the three-dimensional gravitational potential (the so-called Weyl potential \cite{Lewis:2006:review}), $\Psi$, sourced by the matter distribution, as
\al{
    \phi(\hatn) =  -2 \int_0^{\chi^*} d\chi \frac{\chi^*-\chi}{\chi^*\chi} \Psi(\chi \hatn, \eta_0 - \chi)
    \,. \label{eq:mass_field}
}
Here, $\chi$ is the conformal distance, $\eta_0$ is the conformal time today, and $\chi^*$ is the conformal distance to the last-scattering surface. \footnote{The temperature and polarization anisotropies are generated at slightly different epochs and durations. Since the lensing kernel is almost insensitive to a slight change in $\chi^*$ \cite{Hadzhiyska:2017nqe}, we assume $\chi^*$ is the same for the temperature and polarization.} 
The gravitational potential, $\Psi(\chi \hatn, \eta_0-\chi)$, is evaluated along the unperturbed trajectory $\chi\hatn$, the so-called Born approximation \cite{Pratten:2016,Lewis:Pratten:2016,Fabbian:2017wfp}, with conformal time $\eta_0-\chi$. We ignore the curl mode, which vanishes in the linear theory of perturbations having only scalar density perturbations (see, e.g., Ref.~\cite{Namikawa:2014:review} and reference therein). 

Furthermore, the lensing converts part of the $E$-mode polarization into $B$-mode polarization \cite{Zaldarriaga:1998ar}. The lensing-induced $B$-mode polarization is roughly comparable to a $5\,\mu$K-arcmin white noise up to half-degree angular scale and has been detected by ground-based CMB experiments (e.g., Ref.~\cite{Hanson:2013:lensB}). This lensing $B$-mode polarization impedes the detection of inflationary gravitational waves \cite{Seljak:2003pn}. \paperII\ focuses on overcoming this issue with delensing techniques, which we will also mention in Sec.~\ref{sec:applications}.

\subsection{Reconstruction of Lensing Potential}

The lensing potential can be reconstructed using the fact that averaging over CMB realizations, while keeping the lensing potential unchanged, violates the statistical isotropy of the CMB, thereby introducing correlations between different CMB polarization multipoles. The off-diagonal elements of the CMB polarization covariance generated by lensing up to linear order in $\phi$ are given as \cite{OkamotoHu:quad,Lewis:2011:bispec} 
\footnote{
Cosmic birefringence -- a rotation of the CMB linear polarization plane as they travel through space (see Ref.~\cite{Komatsu:2022:review} and references therein) can also lead to a nonzero off-diagonal element. However, the additional contributions do not bias the lensing estimate due to the difference in the parity symmetry \cite{Namikawa:2021gbr}. 
}
\begin{equation}
    \langle X_{\ell m} Y_{\ell'm'}\rangle_{(\l m)\not=(\l',-m')} = \sum_{LM} (-1)^M  \left(\begin{array}{clcr}
    \ell & \ell' & L\\
    m & m' & -M  \end{array}\right) f^{XY}_{\ell \ell' L} \phi_{LM}
    \,, 
\end{equation}
where $X$, $Y\in \{E,B\}$ and the quantity in parentheses is the Wigner 3$j$ symbol that represents the angular momentum coupling. The function $f^{XY}_{\ell \ell' L}$ quantifies the response of the off-diagonal correlations to lensing whose explicit form is given by Table 1 in Ref. \cite{OkamotoHu:quad}. Thus, the lensing potential can be estimated as a quadratic combination of the CMB anisotropies at different angular scales.

In the \LB\ case, the $EB$ estimator dominates the SNR of lensing. The improvement of the SNR using the iterative estimator is negligible for the \LB\ case. Therefore, this paper only focuses on the $EB$ quadratic estimator. In an idealistic case, the estimator is given as \cite{OkamotoHu:quad} 
\al{
    \hat{\phi}_{LM} = A_L^{\phi}(\bar{\phi}_{LM}-\ave{\bar{\phi}_{LM}}) \,,\label{eq:phi_w_mf}
}
where we define 
\al{
    \bar{\phi}^*_{LM} = \sum_{\ell\ell'}\Wjm{\ell}{\ell'}{L}{m}{m'}{M} (f^{EB}_{\ell \ell' L})^* \bar{E}_{\ell m}\bar{B}_{\ell'm'} \,.  
}
In an idealistic case, the filtered multipoles, $\bar{E}_{\ell m}$ and $\bar{B}_{\ell m}$, are obtained by multiplying their inverse variance for each multipole, $\bar{X}_{\ell m}=X_{\ell m}/C^{XX}_{\ell}$, with $X=E$ or $B$. In this study, we employ Wiener filtering, including the pixel-space noise covariance, the details of which are described in Sec.~\ref{sec:cinv_theory}. The normalization in the idealistic full-sky case is given by
\al{
    A_L^{\phi} = \left\{\frac{1}{2L+1}\sum_{\ell\ell'}\frac{|f^{EB}_{\ell \ell' L}|^2}{C^{EE}_{\ell}C^{BB}_{\ell'}}\right\}^{-1} 
    \,.
}
In the following, we describe how to implement the estimator for \LB.

\subsection{Galactic Foreground Cleaning}

We perform lensing reconstruction on the foreground-cleaned polarization map. Our simulation includes both Galactic dust and synchrotron emission, with spatially varying spectral indices. Point sources and free-free emission are excluded, since their impact on large angular CMB multipoles is insignificant \cite{Smith_PS,macellari_FF}. The harmonic coefficients of our frequency maps at each observed frequency $\nu$ are modeled as
\begin{equation}
    M_{\ell m}^\nu = S^{\rm CMB}_{\ell m} + S_{\ell m}^{{\rm FG},\nu} + N_{\ell m}^\nu
    \,,
\end{equation}
where $S^{\rm CMB}_{\ell m}$ is the CMB component, $S_{\ell m}^{{\rm FG},\nu}$ is the total foreground contribution at the observing frequency $\nu$, and $N^\nu_{\ell m}$ is the noise at the frequency $\nu$. 
The assumption of a known frequency scaling permits the utilization of ILC in harmonic space for the cleaning. An additional critical assumption underlying the ILC method is the Gaussianity of the CMB. In HILC, we define the foreground-cleaned map as~\cite{Tegmark:2003}
\begin{equation}
    S^{\rm cleaned}_{\ell m} = \bm{w}_{\ell} \cdot \bm{M}_{\ell m} \,,
\end{equation}
where $\bm{w}_{\ell}$ contains the weights for each frequency map, and the bold letters indicate the vectors containing all observed frequency maps. We can derive the HILC weights by minimizing the variance from the foregrounds and noise contributions under the constraint for an unbiased estimate, $\bm{w}_\ell^\mathrm{T} \cdot \bm{a} = 1$, with $\bm{a}=(1,1,\dots,1)^\mathrm{T}$. The weights are then given by
\begin{equation}
    \bm{w}_{\ell} = \frac{\bR{C}_{\ell}^{-1} \bm{a}}{\bm{a}^\mathrm{T} \bR{C}_{\ell}^{-1} \bm{a}}
    \,,
\end{equation}
where the covariance of frequency maps in harmonic space is given by
\begin{equation}
    \bR{C}_{\ell} = \frac{1}{2\ell+1} \sum_m \bm{M}_{\ell m} \bm{M}_{\ell m}^\dagger
    \,. 
\end{equation}
The HILC method requires the total power spectrum, which includes contributions from the CMB, Galactic foregrounds, and noise, to compute the covariance of the frequency maps. In practice, this total power spectrum is estimated from observed data. On the other hand, the method does not include the non-Gaussianity of the Galactic foregrounds. 
Unlike other component-separation methods, the HILC method focuses on minimizing the variance in the foreground cleaned map while preserving the signal power, rather than globally suppressing the statistical noise.
HILC assumes isotropy to derive the weights, in contrast to, for example, Needlet ILC \cite{Delabrouille:2008:NILC}. This isotropic assumption limits the ability of HILC to optimally account for spatial variations in foreground morphology and spectral energy distribution (SED), potentially introducing bias in the lensing measurements due to the spatial variation of Galactic foregrounds. This paper tests whether the HILC method works even if the Galactic foregrounds have some spatial variation. 

\subsection{Filtering of CMB anisotropies}\label{sec:cinv_theory}

We compute the Wiener-filtered $E$- and $B$-mode polarization starting from the above component-separated polarization map as inputs. Specifically, we solve the following equation~\cite{Eriksen:2004:wiener,Namikawa:2021:SO-delens}: 
\al{
	&\left[1+\bR{C}_{\rm s}^{1/2}\bR{B}\bR{Y}_2^\dagger\bR{N}_{\rm pix}^{-1}\bR{Y}_2\bR{B}\bR{C}_{\rm s}^{1/2}\right] (\bR{C}_{\rm s}^{-1/2}\bm{\hpol}^{\rm WF}) 
	= \bR{C}^{1/2}_{\rm s}\bR{B}\bR{Y}^\dagger_2 \bR{N}_{\rm pix}^{-1} \bm{\hpol}^{\rm obs}
	\,. \label{Eq:Cinv}
}
Here, we solve for the vector $\bm{\hpol}^{\rm WF}$, which has the harmonic coefficients of the Wiener-filtered $E$- and $B$-mode polarization, $\bR{C}_{\rm s}$ is the diagonal signal covariance of the lensed $E$- and $B$-mode polarization in spherical-harmonic space, and $\bR{B}$ is a diagonal matrix to include the beam smearing. The matrix $\bR{C}_{\rm s}^{1/2}$ is defined so that its square is equal to $\bR{C}_{\rm s}$. The real-space vector $\bm{\hpol}^{\rm obs}$ contains the observed Stokes $Q$ and $U$ maps after adopting HILC. The matrix $\bR{Y}_2$ is defined to transform the multipoles of the $E$- and $B$-modes into real-space maps of the Stokes parameters $Q$ and $U$. Finally, $\bR{N}_{\rm pix}$ is the pixel-space covariance matrix of the instrumental noise in these maps. 

To compute the noise covariance, we assign infinite noise for pixels not used in the analysis. 
We assume isotropic noise in observed pixels, following a model of $E$- and $B$-mode noise power spectra. Explicitly, we assume that the inverse noise covariance is given by 
\al{
    \bR{N}_{\rm pix}^{-1} = \bR{W}\bR{Y}_2\bR{N}^{-1}\bR{Y}^\dagger_2\bR{W}
    \,, 
}
where $\bR{W}$ is a matrix that takes the value 1 for pixels used in the analysis and zeroes otherwise, and $\bR{N}$ is the noise covariance of $E$ and $B$ modes. We assume that the noise covariance of $E$- and $B$- modes is diagonal in harmonic space, with elements given by a model of the $E$- and $B$- mode noise power spectra, $N^{EE}_\ell$ and $N_\ell^{BB}$. 
Our simulation pipeline uses the noise spectra obtained from the HILC weights during component separation with the reference noise spectra of $15$ frequency channels. We do not include any extra masks besides the Galactic mask.

\section{Simulations} \label{sec:simulation}

\begin{figure}[t]
    \centering
    \includegraphics[width=1\textwidth]{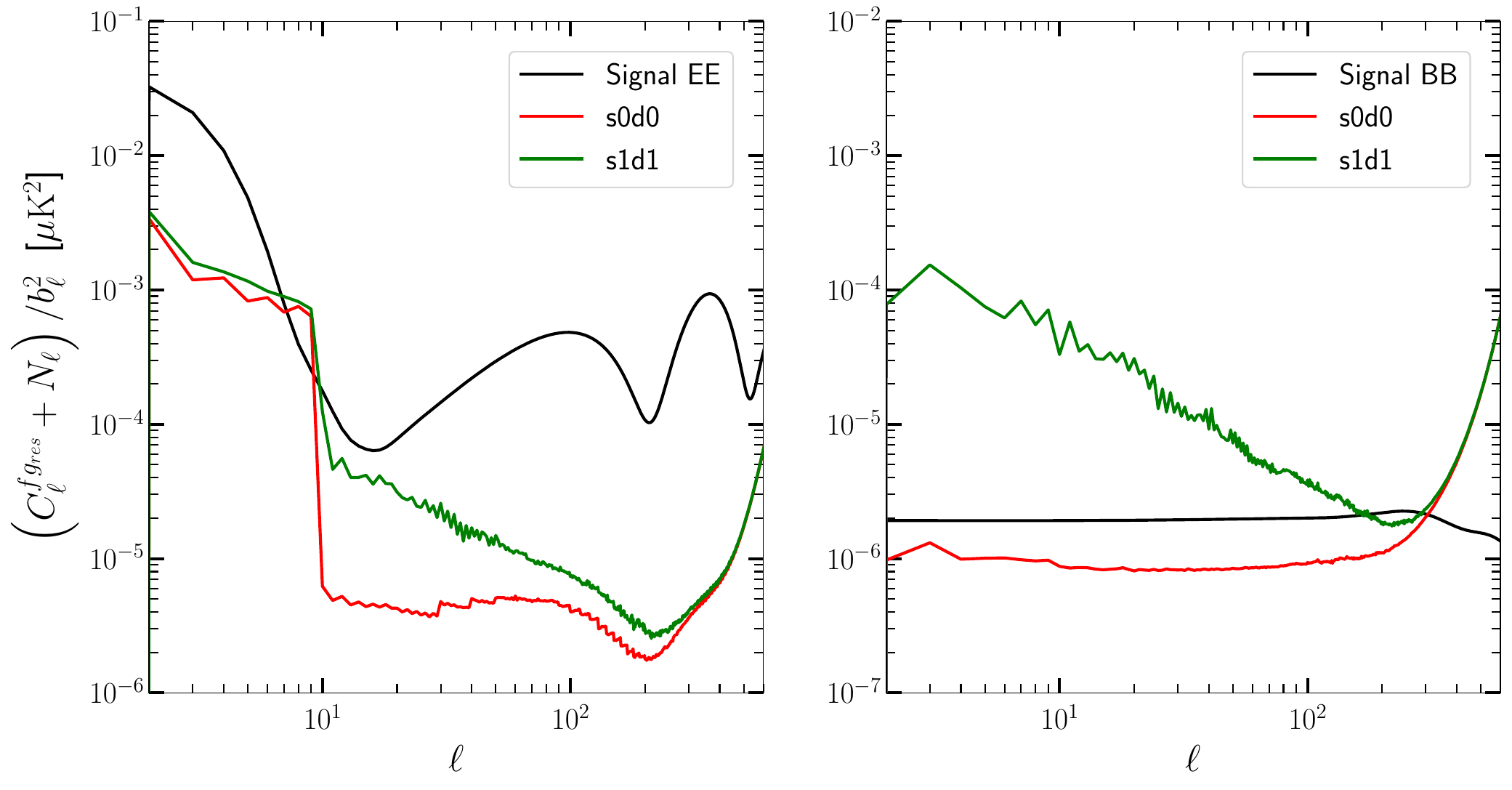}
    \caption{
    Foreground residuals and noise presented in our component-separated polarization maps. The solid black line represents the signal. The left and right panels display the $E$- and $B$-mode residuals, respectively. We illustrate the residuals for two foreground models, \textbf{s0d0} (red) and \textbf{s1d1} (green). Note that the green curves deviate significantly from the assumed model during component separation, substantially increasing residuals. Both spectra are computed from $500$ full-sky simulations.
    }
    \label{fig:cs_spectra}
\end{figure}

This section overviews the simulation sets used in this analysis and in \paperII\ to produce the results. We use the experimental specifications described in table $13$ of \pteppaper\ to generate multiple frequency maps with frequency coverage ranging from $40\,$GHz to $402\,$GHz in $15$ frequency channels. 
Note that while table $13$ of \pteppaper\ provides $22$ frequency channels, we exclusively consider the combined values for each frequency. However, our investigation finds that the changes in the noise properties remain negligible in the foreground-cleaned map, even when utilizing the full $22$-frequency configuration.

We generate realizations of the frequency maps and the post-component-separation map through the following steps.
\begin{enumerate}
\item 
We create a lensed CMB polarization map of the full sky. We first generate unlensed CMB polarization maps and a lensing potential map on the full sky as random Gaussian fields drawn from the input fiducial angular power spectra. We then remap the unlensed CMB maps with the lensing potential map using \textsc{lenspyx},\footnote{\url{https://github.com/carronj/lenspyx}} where the algorithm utilizes bicubic interpolation in an oversampled equidistant-cylindrical-projection grid.
\item 
We produce $15$ frequency maps of synchrotron and dust emission using \textsc{pysm3}.\footnote{\url{https://github.com/galsci/pysm}} For our baseline simulation set we use \textbf{s1} and \textbf{d1} models for synchrotron and dust, respectively. In the \textbf{s1} model, a power law scaling is applied to the synchrotron emission templates \cite{Remazeilles:2014:s1model,Bennett:2013:wmap} with a spatially varying spectral index \cite{Miville-Deschenes:2008:FG}. The thermal dust, \textbf{d1}, is modeled as a single-component modified blackbody. The \Planck\ dust template is scaled to different frequencies with a modified-blackbody spectrum using spatially varying temperature and spectral index \cite{P15:FG}. 
In our analysis, we also consider the \textbf{d0} model for dust and the \textbf{s0} model for synchrotron radiation, both sourced from PySM. Here, the \textbf{d0} model corresponds to a simplified version of the \textbf{d1} model, characterized by a fixed spectral index of $1.54$ and a blackbody temperature of $20\,$K. On the other hand, the \textbf{s0} has a constant spectral index of $-3$.
\item 
Each frequency map containing the CMB signal and the foreground components is convolved with the associated Gaussian beam, and then the corresponding white noise is added to each frequency map.
\item 
We deconvolve the aforementioned beam-convolved frequency maps and perform component separation using the HILC algorithm with \textsc{fgbuster}\footnote{\url{https://github.com/fgbuster/fgbuster}} to obtain the full-sky component-separated map. HILC uses an evenly binned covariance, with bin size, $\Delta \ell = 10$.
\end{enumerate}
We repeat this process for each realization and generate $500$ realizations (hereafter SET1). 
These maps are defined on a \textsc{Healpix} \cite{gorski} grid with $N_{\rm side}=512$. In addition, we generate a simulation set to specifically compute the so-called $N^{(1)}$ bias in the lensing measurement \cite{Kesden:2003:N1}, where the input lensing map is kept fixed for all realizations. This fixed-lensing simulation is generated for $100$ realizations (hereafter SET2). 

Figure \ref{fig:cs_spectra} shows the $E$- and $B$-mode angular power spectra of the component-separated CMB maps on the full sky. At large angular scales ($\ell\alt 100$), the residual foregrounds become non-negligible, especially for the $B$-mode power spectrum, which is dominated by residual foregrounds at $\ell\alt 100$. At smaller angular scales ($\ell\agt 600$), the instrumental noise becomes dominant. 

\section{Lensing Reconstruction} \label{sec:results}

In this section, we present the results of the reconstruction of lensing potential from our simulation set. For the forecast, we consider the \textbf{s1d1} foreground model as our baseline simulation. We also highlight the biases in our estimates of the lensing power spectrum.

\subsection{Reconstructed Lensing Map}
We reconstruct the lensing potential from the component-separated maps using the public \Planck\ Galactic mask\footnote{\href{http://pla.esac.esa.int/pla/aio/product-action?MAP.MAP_ID=HFI_Mask_GalPlane-apo0_2048_R2.00.fits}{\text{HFI\_Mask\_GalPlane\-apo0\_2048\_R2.00.fits}}} that keeps $80\%$ of the sky. Before this reconstruction, we apply the filtering as described in Eq.~\eqref{Eq:Cinv} of Sec.~\ref{sec:cinv_theory} to the component-separated maps. The HILC-weighted noise spectra averaged over the SET1 simulation are used for the noise covariance, $\bR{N}$. We employed the $EB$ quadratic estimator with \textsc{cmblensplus}\footnote{\url{https://github.com/toshiyan/cmblensplus}} \cite{cmblensplus} for reconstructing the lensing deflection field. 
For the analysis of this paper, CMB multipoles $2<\ell<600$ were used. The maximum multipole is determined so that the SNR of the lensing signals is saturated. \footnote{For \paperII, we also compute the lensing reconstruction with CMB multipoles at $190 <\ell <600$ to eliminate potential delensing biases on the delensed power spectrum \cite{Seljak:2003pn}.}
\begin{figure}[t]
    \centering
    \includegraphics[width=1\textwidth]{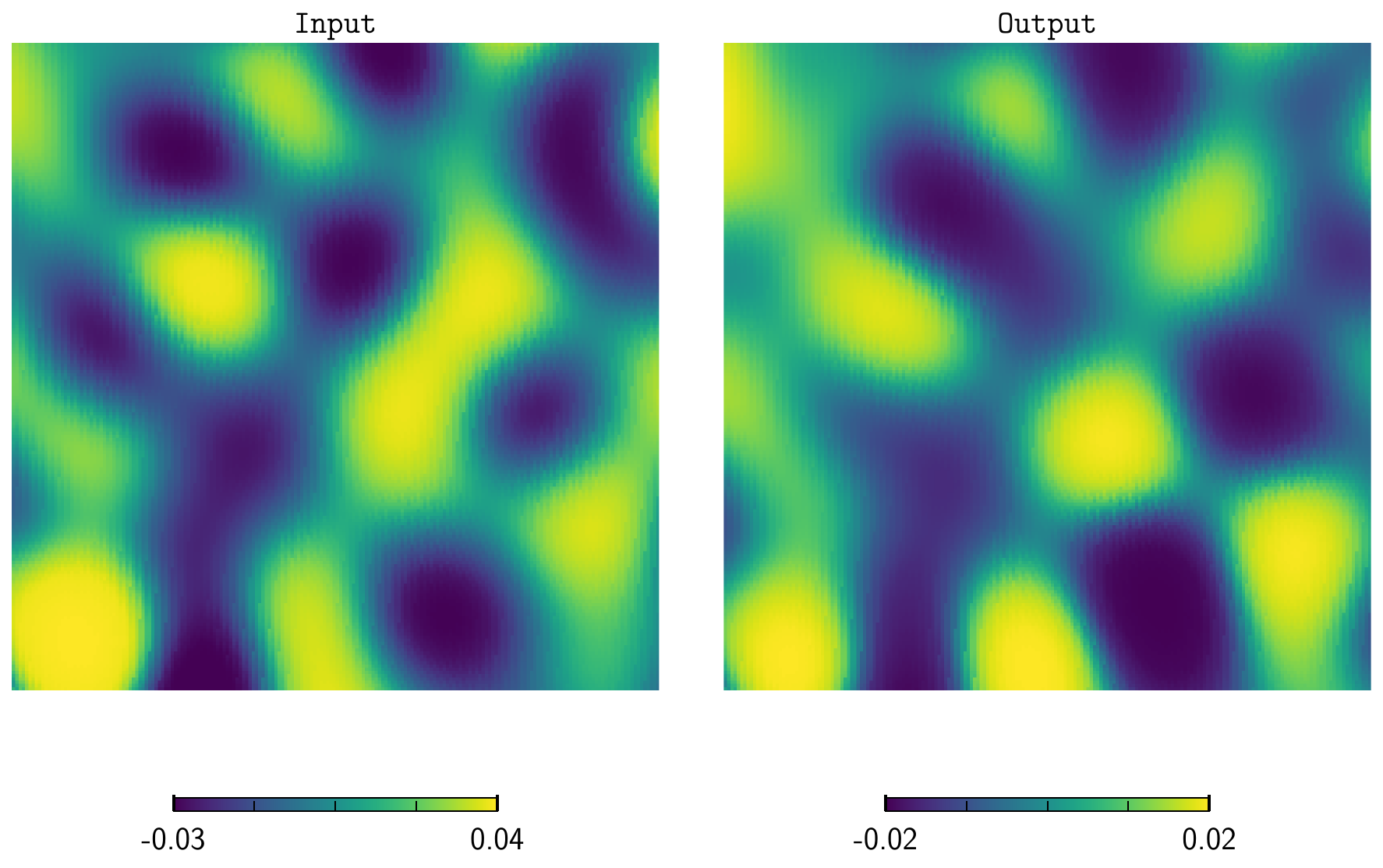}
    \caption{Lensing convergence map, which is defined in harmonic space as $\kappa_{LM}\equiv L(L+1)\phi_{LM}/2$ in harmonic space, obtained from one realization of the SET1 \textbf{s1d1} simulation. The map contains multipole of $2<L<100$. \textit{Left}: The input map used to remap the primary CMB. \textit{Right}: The Wiener-filtered reconstructed map. The map resolution is $5$ arcmin-pixel$^{-1}$.}
    \label{fig:deflection}
\end{figure}
Figure \ref{fig:deflection} shows one realization of the reconstructed lensing-convergence map. We can see a clear correlation between the input and reconstructed lensing maps.

\subsection{Biases in the Lensing Power Spectrum Estimate}

\begin{figure}[t]
    \centering
    \includegraphics[width=.9\textwidth]{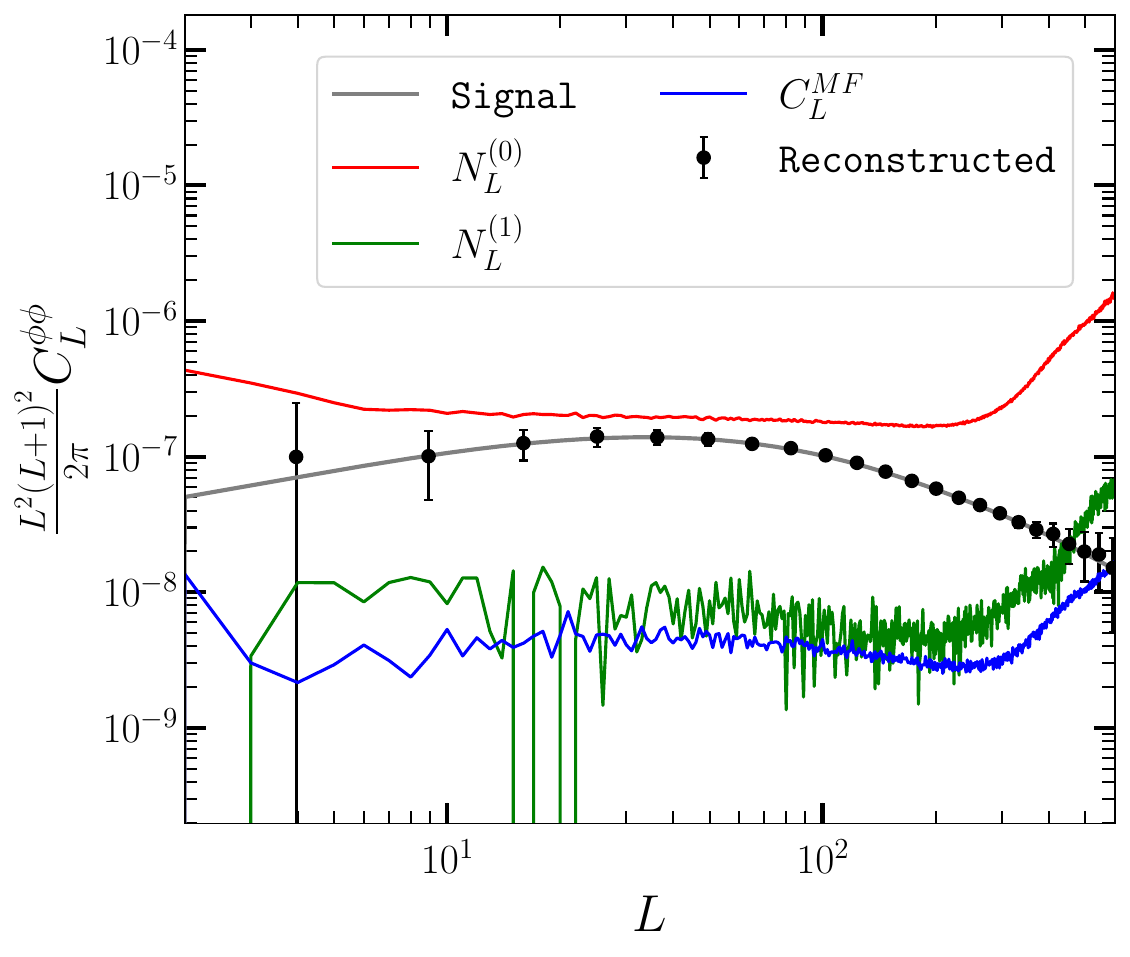}
    \caption{Binned angular power spectrum of the lensing potential shown in the black with error bars. The mean and standard deviation are obtained from $400$ realizations of the SET1 simulation. The red and green solid lines are the MCN0 and MCN1 biases, respectively. The mean field is shown with a blue line. We use the \textbf{s1d1} model for this analysis. We show the power spectrum up to $L=600$ since the power spectrum at $L>600$ is significantly noise-dominant.}
    \label{fig:clpp}
\end{figure}

The estimation of the angular power spectrum of the lensing potential requires the subtraction of several biases because the estimates of the lensing potential are quadratic in CMB polarization anisotropies, and the power spectrum of the lensing estimator is the four-point correlation of the CMB anisotropies. The CMB four-point correlation consists of contributions from disconnected and connected parts, and the latter contains the lensing potential power spectrum. The remaining terms emerge as a bias in estimating the lensing power spectrum \cite{Kesden:2003:N1,Hanson:2010:N2,Boehm:2016,Boehm:2018}. The power spectrum of the quadratic estimator, $C_L^{\hat{\phi}\hat{\phi}}$, is described by
\begin{equation}
    C_L^{\hat{\phi}\hat{\phi}} = R_L C_L^{\phi\phi} + N_L^{(0)} + N_L^{(1)} + \Delta C^{\phi\phi}_L 
    \,.
\end{equation}
The first term is the lensing potential power spectrum, $C_L^{\phi\phi}$, multiplied by a factor, $R_L$, due to a mismatch between the analytic and true normalizations. The second term, $N_L^{(0)}$, arises from the disconnected part of the four-point correlation, the so-called $N0$ bias, which receives contributions from the CMB, foreground, and noise. The third term, $N_L^{(1)}$, is the so-called $N1$ bias that arises from the secondary contraction of the connected parts by lensing at first order in $C_L^{\phi\phi}$ \cite{Kesden:2003:N1}, and the last term, $\Delta C^{\phi\phi}_L$, contains the other remaining biases of $C^{\phi\phi}_L$ due to, e.g., possible issues of the estimation of normalization, mixing of power between different multipoles, and biases at ${\cal O}((C_L^{\phi\phi})^2)$. 

\begin{figure}[t]
    \centering
    \includegraphics[width=.8\textwidth]{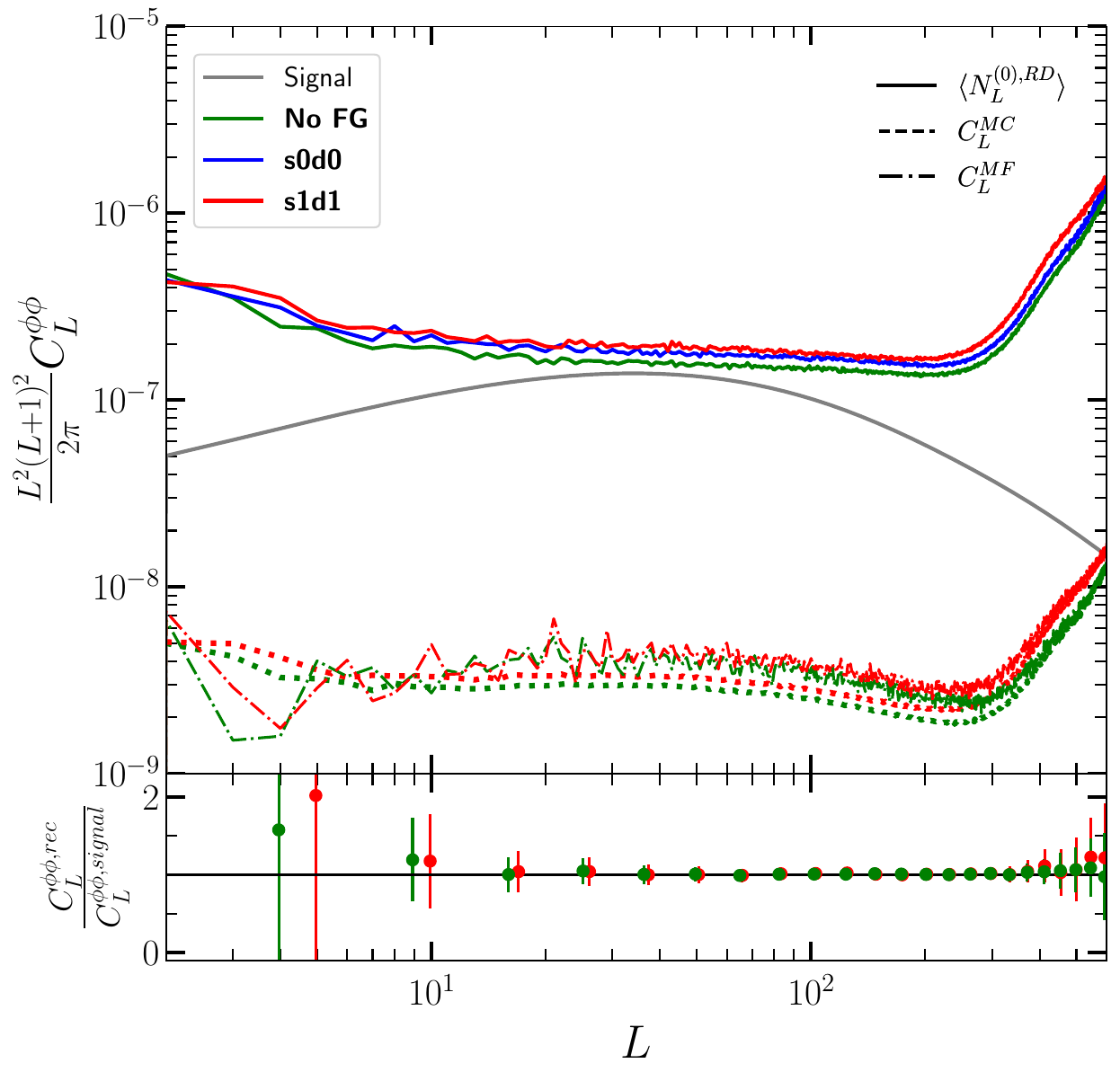}
    \caption{{\it Top}: Comparison plot of the N0 (upper, colored solid) and mean-field biases (lower, colored dash-dotted), with and without the Galactic foregrounds. For the N0 bias, we show two foreground cases, \textbf{s0d0} and \textbf{s1d1}, in addition to the no foreground case (`No FG'). We also show the Monte Carlo error in estimating the mean-field bias, $C_L^{\rm MC}$, (dotted), and the lensing power spectrum signal (gray solid). {\it Bottom}: The ratio of the measured lensing power spectrum to the input lensing power spectrum with (red) and without (green) the Galactic foregrounds. 
    Note that the power spectrum shown in red is shifted to the right as $L\to L+1$ for clarity.
    }
    \label{fig:FG-and-noFG}
\end{figure}

We estimate the mean-field bias, $\ave{\hat{\phi}_{LM}}$, which arises primarily due to the mask and foreground residuals, using $100$ realizations of the SET1 simulation. The mean-field bias is subtracted at the map level as shown in Eq.~\eqref{eq:phi_w_mf}, and these realizations are not included in computing other bias terms. To check the level of the mean-field bias, we also compute the power spectrum of the mean-field bias, $C_L^{\rm MF}$. 

The $N0$ bias approximately corresponds to the noise power spectrum of the reconstructed lensing map. We compute the $N0$ bias with a Monte Carlo simulation (hereafter MCN0). Specifically, we compute the lensing estimator using the SET1 simulation in the following way: 
\begin{equation}
    N_L^{(0),\rm MC} \equiv \Big\langle 
        C_L^{\hat{\phi}\hat{\phi}}\left[E^i,B^j,E^i,B^j \right] 
        +
        C_L^{\hat{\phi}\hat{\phi}}\left[E^i,B^j,E^j,B^i \right]
         \Big \rangle_{\rm SET1}
    \,, \label{eq:mcn0}
\end{equation}
where $i$ and $j$ are simulation indices, and we define
\al{
    C_L^{\hat{\phi}\hat{\phi}}\left[X,Y,X',Y'\right] \equiv C_L^{\hat{\phi}^{XY}\hat{\phi}^{X'Y'}}
    \,. 
}
The ensemble average is over $500$ realizations from SET1 with $i$ and $j=i+1$ changing cyclically. Since the MCN0 estimate is not optimal \cite{Schmittfull:2013,Peloton:2016:RDN0}, we also compute the realization-dependent $N0$ (hereafter RDN0) bias \cite{namikawa-MF,P15:phi} for each realization in SET2, which is given by
\begin{equation}\label{eq:rdn0}
\begin{split}
    N_{L,i}^{(0),\rm RD} &\equiv \langle 
        C_L^{\hat{\phi}\hat{\phi}}\left[E^i,B^j,E^i,B^j \right] 
        +
        C_L^{\hat{\phi}\hat{\phi}}\left[E^j,B^i,E^i,B^j \right] 
        \\
        &\quad +
        C_L^{\hat{\phi}\hat{\phi}}\left[E^i,B^j,E^j,B^i \right]
        +
        C_L^{\hat{\phi}\hat{\phi}}\left[E^j,B^i,E^j,B^i \right] 
        \rangle_{j}
        \\
        &\quad - N_{L}^{(0),\rm MC}\,.
\end{split}
\end{equation}
The average is over $100$ realizations with $j=i+1$, which is independent of that used for evaluating $N_L^{(0),\rm MC}$. The $N1$ bias is estimated with the SET2 simulation as (e.g., Ref.~\cite{Story:2015}, hereafter MCN1)
\begin{equation} 
     N_L^{(1),\rm MC} \equiv \left\langle
        C_L^{\hat{\phi}\hat{\phi}}\left[E^i,B^j,E^i,B^j\right] 
        +
        C_L^{\hat{\phi}\hat{\phi}}\left[E^i,B^j,E^j,B^i\right] \right\rangle_\text{SET2}
        - N_L^{(0),\rm MC}        
    \,. \label{eq:mcn1}
\end{equation}
Here, we use $100$ realizations of the SET2 simulations to evaluate the ensemble average. 

Figure \ref{fig:clpp} shows the significance of each bias term compared with the input theoretical lensing power spectrum. We also show the debiased angular power spectrum of lensing potential:
\begin{equation}\label{eq:cl_pp_final}
    \hat{C}_L^{\phi\phi} = \frac{1}{R^{\rm MC}_L}\left(C_L^{\hat{\phi}\hat{\phi}} - N_L^{(0),\rm RD} - N_L^{(1),\rm MC}\right)
    \,,
\end{equation}
where $R^{\rm MC}_L$ is the normalization correction obtained from
\begin{equation}
    R^{\rm MC}_L = \left(\frac{C_L^{\phi^{\rm in}\hat{\phi}}}{C_L^{\phi^{\rm in}\phi^{\rm in}}}\right)^2 
    \,.
\end{equation}
Here, $\phi^{\rm in}$ is the full sky input lensing map. 
We use $400$ realizations of the SET1 simulations to obtain the mean and standard deviation of $\hat{C}_L^{\phi\phi}$. 
After correcting for the N0 bias, N1 bias, and normalization, the power spectrum is in good agreement with the input power spectrum. This indicates that $\Delta C^{\phi\phi}_L$ is negligible. The result also shows that the foreground-induced trispectrum and mean-field bias are negligible. 

In Figure \ref{fig:FG-and-noFG}, we compare the $N0$ and mean field biases of the lensing reconstruction in the presence and absence of Galactic foregrounds. Without Galactic foregrounds, the only non-idealistic effect is the Galactic mask, which primarily causes the mean-field bias. With the foreground cleaning, the noise level of the cleaned CMB polarization map increases, leading to an increase in the $N0$ bias. 
The power spectrum of the mean-field bias is more than an order of magnitude lower than the signal power spectrum. The mean field does not increase even in the presence of the residual Galactic foregrounds. 

\subsection{The signal-to-noise ratio}

\begin{figure}[t]
    \centering
    \includegraphics[scale=.5]{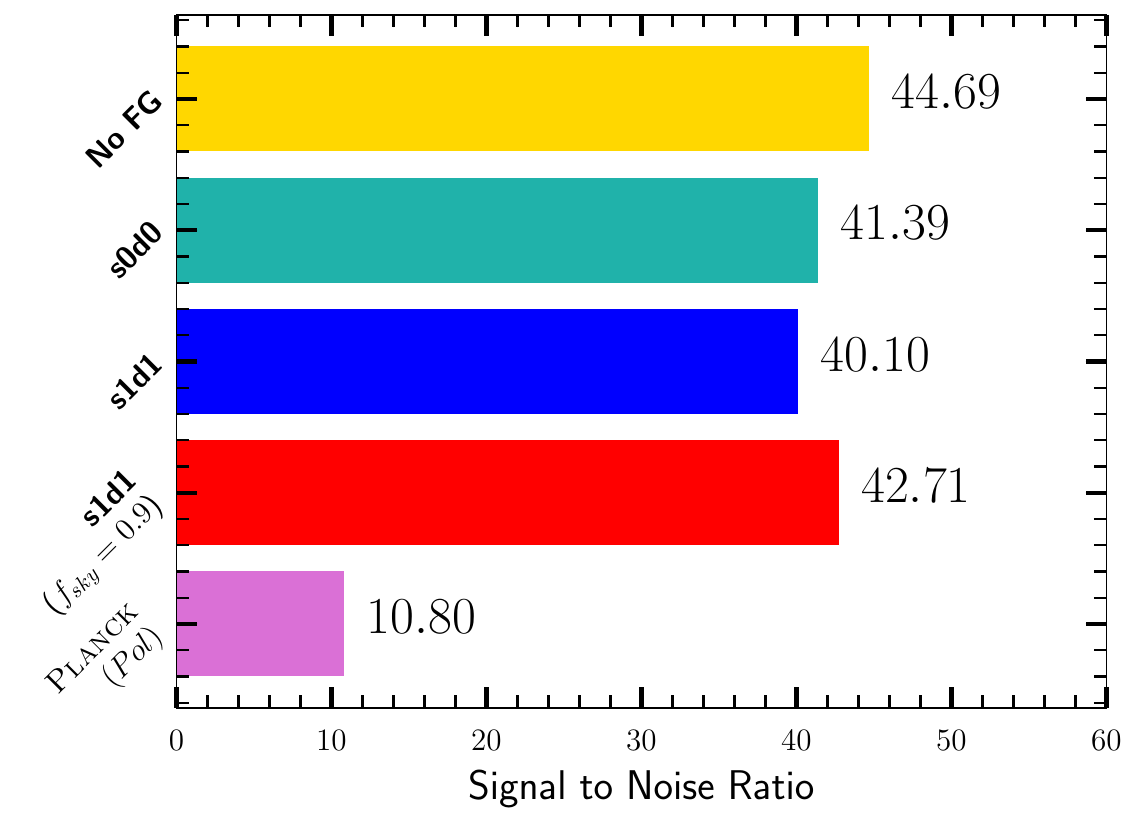}
    \caption{SNR of the lensing power spectrum. The yellow bar shows the case with no foregrounds, while the light green and blue bars represent the cases with the \textbf{s0d0} and \textbf{s1d1} foreground models, respectively. The red bar shows the case with the \textbf{s1d1} foreground model but with a wider sky coverage, $f_{\rm sky}=0.9$. The pink bar is the SNR from the \Planck\ polarization-based lensing analysis~\cite{Carron:2022:NPIPE-lensing}. 
    }
    \label{fig:snr}
\end{figure}

We here show a simulation-based estimate of the SNR of the lensing signal. We first measure the amplitude of the lensing spectrum from a weighted mean over multipole bins:
\begin{align} 
    \hat{A}_{\rm lens} = \frac{\sum_b a_b \hat{A}_b}{\sum_b a_b} 
    \,. \label{Eq:AL}
\end{align}
The quantity $\hat{A}_b$ is the relative amplitude of the power spectrum compared with a fiducial power spectrum for the \Planck\ $\Lambda$CDM cosmology, $C_b^{\phi\phi}$, i.e., $\hat{A}_b \equiv \hat{C}_b^{\phi\phi}/C_b^{\phi\phi}$. The weights, $a_b$, are taken from the band-power covariance as
\begin{align} 
    a_b = \sum_{b'} C^{\phi\phi}_b {\bm {\mathrm{Cov}}} _{bb'}^{-1} C^{\phi\phi}_{b'} 
    \,.
\end{align}
The fiducial band-power values and their covariances, including off-diagonal correlations between different multipole bins, are evaluated from the simulations. The SNR is then given by
\begin{equation}
    {\rm SNR} = \frac{1}{\sigma_A}
    \,,
\end{equation}
where $\sigma_A$ is the $1\,\sigma$ constraint on $A_{\rm lens}$ computed from SET1 simulations. 

Figure \ref{fig:snr} shows a summary of the SNR for three cases: the baseline, \textbf{s0d0}, and no-foreground cases. We find that the SNR of the \LB\ lensing reconstruction is comparable to that of the latest full-sky lensing measurement from \Planck\ \cite{P18:phi}. In the presence of the \textbf{s0d0} foregrounds, the SNR decreases by $10$\% compared to the no-foreground case. If we consider the foregrounds with the spatially varying spectral index, the SNR decreases by $3$\% compared to the \textbf{s0d0} case. It is worth noting that \LB\ uses polarization alone, and the reconstructed map from \LB\ is complementary to the \Planck\ lensing map, which is mostly based on temperature anisotropies. If we increase the sky fraction to $f_{\text{sky}} = 0.9$, the SNR increases by $6.8$\%, a factor of 4 improvement compared to the \Planck\ polarization-only reconstruction (referred to as Pol hereafter). 

\begin{figure}[t]
    \centering
    \includegraphics[scale=.5]{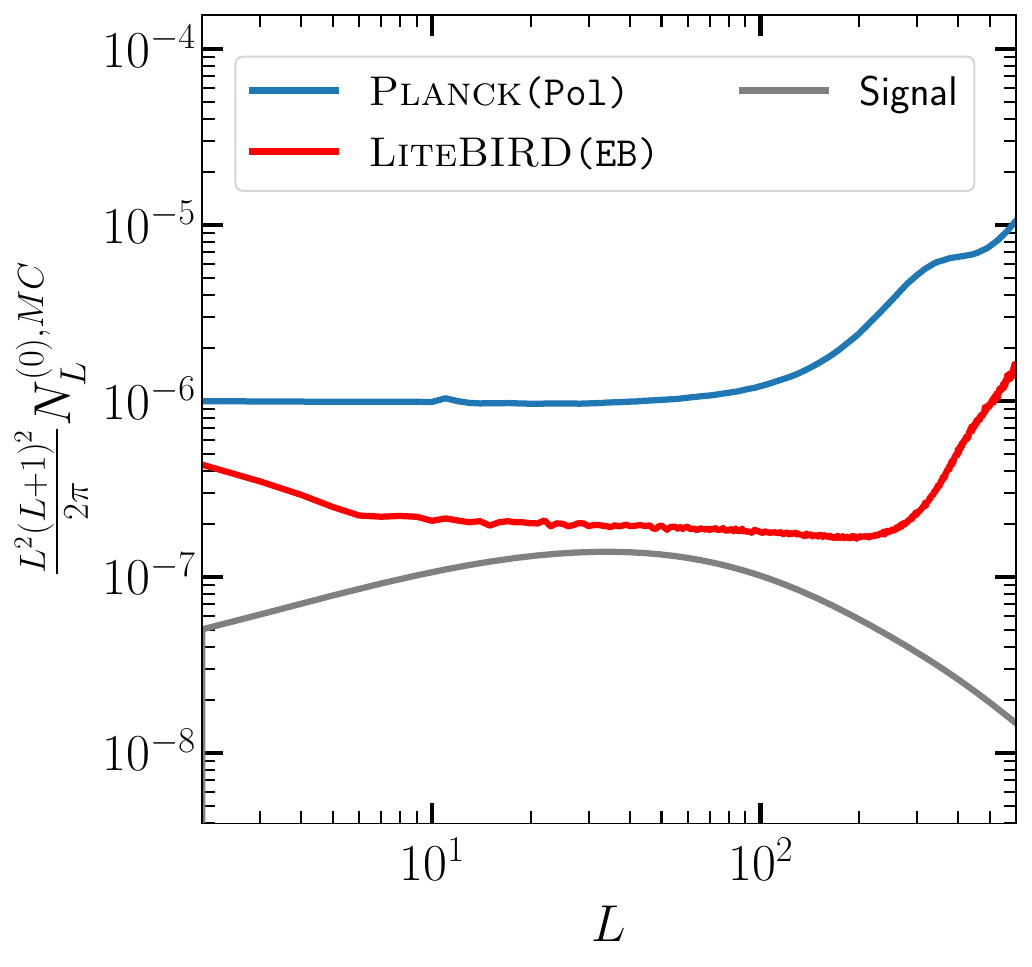}
    \caption{
    Comparison of $N_L^{(0),{\rm MC}}$ between \Planck\ and \LB. The gray solid line represents the signal, while the blue and red solid lines correspond to the noise bias in the \Planck\ and \LB\ polarization-based estimates, respectively. 
    }
    \label{fig:PvsLB}
\end{figure}

Figure \ref{fig:PvsLB} shows a comparison between $N^{(0),{\rm MC}}_L$ of the \Planck\ polarization-only reconstruction and of \LB. The lensing maps from \Planck\ full-data and \LB\ are reconstructed mostly from temperature and from polarization, respectively, and are almost statistically independent. Thus, the combined lensing map from \Planck\ and \LB\ will have an SNR of approximately $60$. We will show a more accurate estimate of the SNR for the combined lensing map in our future work.

\subsection{Impact of Foregrounds in Lensing estimate}

In this section, we quantify the impact of the foregrounds on the lensing estimate, specifically, by computing the bias in the estimation of $A_\mathrm{lens}$. We assume that the \textbf{s1d1} simulation represents real data. We measure the lensing power spectrum from \textbf{s1d1} and estimate $A_{\rm lens}$ from Eq.~\eqref{Eq:AL}. We also repeat the same analysis but with a simulation generated from an incorrect model, i.e., the no-foregrounds and \textbf{s0d0} cases. The incorrect simulation is used to evaluate the bias terms in the lensing power spectrum estimate. We then compute the covariance from the incorrect simulation and obtain $A_{\rm lens}$ from the measured power spectrum. 


\begin{figure}[t]
    \centering
    \includegraphics[width=.5\textwidth]{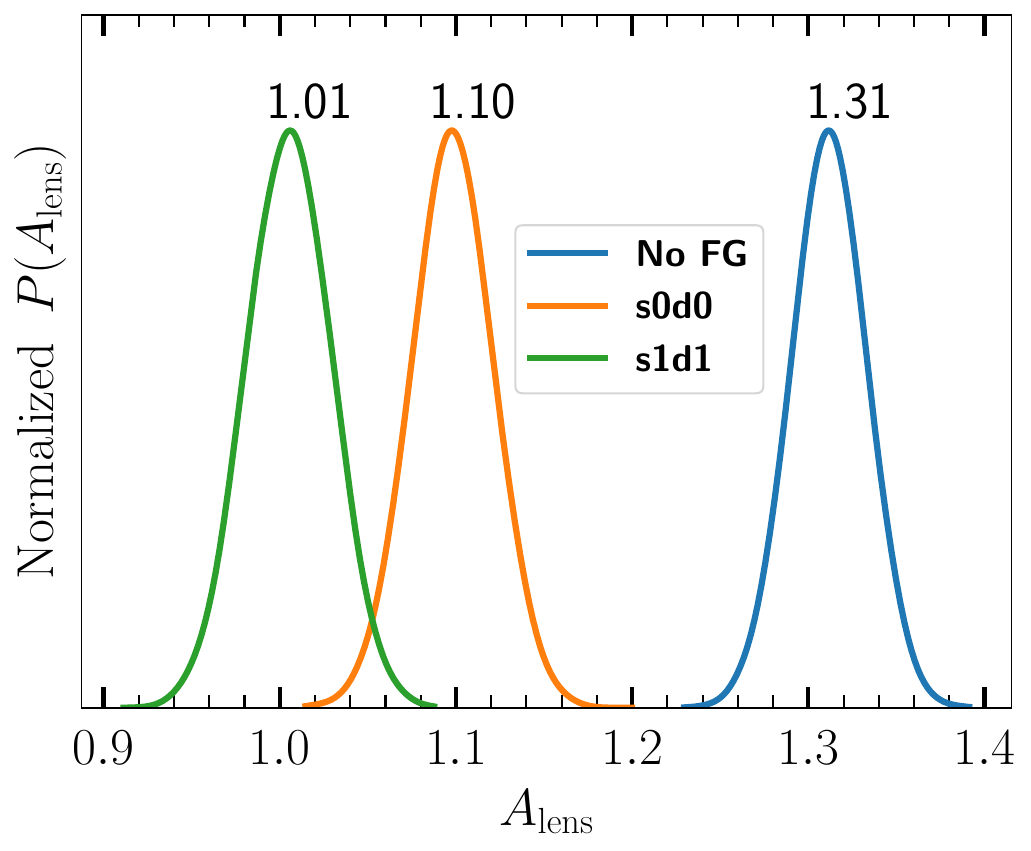}
    \caption{
    Posterior distribution of $A_\mathrm{lens}$ when estimating it for realizations of the \textbf{s1d1} simulation. The blue curve represents the bias in $A_\mathrm{lens}$ when employing the no foreground simulation set to fit the selected spectra. Similarly, the orange and green posterior distributions represent cases where the \textbf{s0d0} and \textbf{s1d1} models were used, respectively. The estimated $A_\mathrm{lens}$ for each case are quoted at the respective peak of posterior.
    }
    \label{fig:alens_bias}
\end{figure}

Figure \ref{fig:alens_bias} presents the posterior distributions of $A_\mathrm{lens}$ assuming that \textbf{s1d1} is real data. The blue posterior corresponds to the estimation of $A_\mathrm{lens}$ when assuming no foregrounds, resulting in a bias of $30$\%. This bias reduces to $10$\% when we utilize the \textbf{s0d0} model, but is larger than the $1\sigma$ statistical error. This is because the two models, \textbf{s0d0} and \textbf{s1d1}, have significant discrepancies already in the $EE$ and $BB$ power spectra, as we show in Fig.~\ref{fig:cs_spectra}. This discrepancy leads to a significant mismatch between the true and simulation disconnected bias. In practice, however, we would not use no-foreground or \textbf{s0d0} for our model since it does not fit the data of $EE$ and $BB$ power spectra. Thus, the bias shown here is more enhanced than what would be obtained in an analysis of real data. We finally mention when we correctly use the simulation from the \textbf{s1d1} model. The bias in $A_\mathrm{lens}$ is negligible compared to $\sigma(A_{\rm lens})$, indicating that potential biases other than those considered in the power spectrum estimation are negligible.\\


\begin{figure}[t]
    \centering
    \includegraphics[width=.5\textwidth]{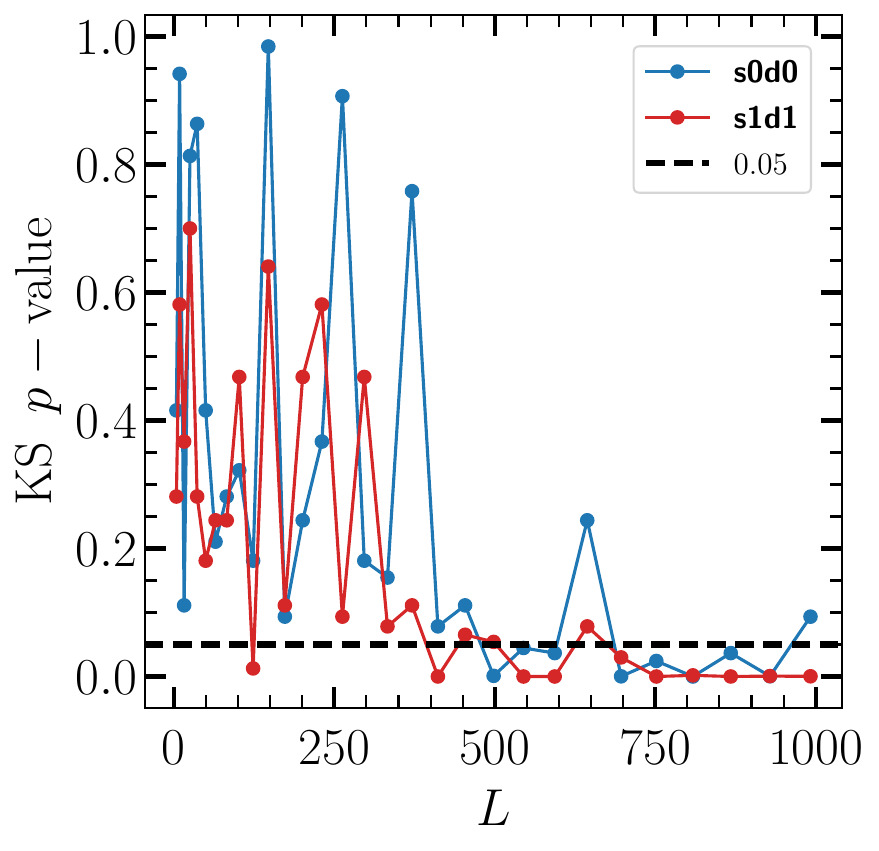}
    \caption{The Kolmogorov-Smirnov test $p$-values obtained for each bin. The red and blue lines represent the $p$-values when comparing \textbf{s0d0} and \textbf{s1d1} with the no-foreground case.}
    \label{fig:pvalue}
\end{figure}

To investigate the agreement of the estimated lensing potential angular power spectra between our simulation sets and detect potential biases in our analysis, we apply the Kolmogorov-Smirnov (KS) test to compute $p$-values within each multipole bin of the angular power spectra. The $p$-values are calculated under the null hypothesis that the power spectra for both no-foreground and foreground cases follow the same distribution after bias mitigation (see Eq.~\ref{eq:cl_pp_final}). The $p$-values for both foreground cases, \textbf{s0d0} and \textbf{s1d1} are shown in Fig \ref{fig:pvalue}. For both foreground cases, the $p$-values for angular scales below $L=400$ exceeded the $0.05$ threshold, indicating a consistent distribution with the no-foreground case. However, for angular scales beyond $L=400$, the $p$-values fell below $0.05$, suggesting a significant deviation. This deviation can be attributed to the non-negligible disconnected bias present in the foreground cases. Despite the discrepancies at high $L$, our analysis demonstrates the effective recovery of the lensing power spectrum, particularly within the angular scales of $L \leq 400$, where the lensing signal dominates.

\section{Applications of LiteBIRD Lensing Map} \label{sec:applications}

In this Section, we discuss some of the potential applications of the \LB\ lensing map. 

\subsection{Cross-correlations}

One potential application of the \LB\ CMB lensing map is to cross-correlate it with other tracers of the large-scale structures in the Universe. The full-sky \LB\ lensing map can be used to calibrate tracers of the large-scale structure through cross-correlation and hence to constrain cosmology. Here, to estimate the SNR of the cross-correlation signal, we use the following equation:
\begin{equation}
    ({\rm SNR})^2 = \sum_{L=L_{\rm min}}^{L_{\rm max}} \frac{f_{\rm sky}^{XY}(2L+1)\left(C^{XY}_L\right)^2}{\left(C^{XY}_L\right)^2 + \left(C^{XX}_L+N_L^{XX}\right)\left(C^{YY}_L+N_L^{YY}\right)}
    \,,
\end{equation}
where $X$ and $Y$ represent the observables as specified below, $N_L^{XX}$ and $N_L^{YY}$ are the associated noise power spectra, and $f_{\rm sky}^{XY}$ is the sky fraction of the overlapping patch of the two observables. 
In this work, we use the baseline noise model, obtained with the \textbf{s1d1} simulations (as described in Section~\ref{sec:simulation}). We do not consider potential biases arising from the bispectrum of the lensing potential, which only leads to a sub-percent level of bias using the $EB$ estimator, even for high-resolution experiments, as shown in Ref.~\cite{Fabbian:2019:cross}.

\subsubsection{Galaxy-lensing cross-correlation as a probe of primordial non-Gaussianity}

The fluctuations of the galaxy number density trace the underlying matter distribution of the large-scale structure and efficiently correlate with the CMB lensing maps. We estimate the SNR of the galaxy-CMB lensing cross-correlation signal ($C^{\phi {\rm g}}_L$) for two different galaxy surveys: the one provided by \Euclid\ \cite{EuclidForecastIST:2020} and the one from the Vera C. Rubin Observatory’s Legacy Survey of Space and Time (LSST, \cite{LSSTCollaboration:2018}). To compute the total SNR, we consider a range of multipoles $L=[2,1000]$. For the \Euclid\ survey, we consider ten equipopulated tomographic galaxy bins and $f_{\rm sky}^{\phi {\rm g}}=0.36$, obtaining ${\rm SNR}=73$. 
For the LSST survey, we consider ten equispaced redshift bins and $f_{\rm sky}^{\phi {\rm g}}=0.35$, obtaining ${\rm SNR}=56$. 
Through the cross-correlation analysis between CMB lensing and galaxy distributions, we can constrain the local primordial non-Gaussianity $f_{\rm NL}$, which induces a scale-dependent bias due to the coupling between long and short wavelength modes (e.g., Refs.~\cite{Dalal:2007:fNL,Jeong:2009:fNL,Schmittfull:2017:fNL,Ballardini:2019:fNL}). To determine the uncertainty of $f_{\rm NL}$ using the cross-spectrum $C^{\phi {\rm g}}_L$ alone, we perform a $\chi^2$ analysis in which we let only $f_{\rm NL}$ free to vary. When considering the cross-correlation between \Euclid\ galaxies and \LB\ CMB lensing, the resulting constraints on $f_{\rm NL}$ are $\sigma(f_{\rm NL})=44$. Note that in this analysis, we vary only $f_{\rm NL}$, and therefore, we do not explore potential degeneracies with other cosmological parameters.

\subsubsection{CIB-lensing cross-correlation}

The cosmic infrared background (CIB) is the integrated emission from unresolved dusty star-forming galaxies. Produced by the stellar-heated dust within galaxies, the CIB carries a wealth of information about the star formation process. 
The CIB traces the matter distribution at a relatively high redshift compared to galaxies in typical optical redshift surveys and is strongly correlated with CMB lensing. This cross-correlation can be used to constrain CIB models \cite{McCarthy:2020:cib} and cosmology \cite{McCarthy:2022:fNL}. 
\Planck\ has measured the CIB-lensing cross-correlations with an SNR of $40$ (statistical uncertainties only) \cite{P13:CIBxphi}. We estimate the SNR of the CIB-lensing cross-correlation, assuming the CIB anisotropies measured by \Planck\, which utilizes the model of CIB described in Appendix D of Ref.~\cite{P15:phi}. We choose $L_{\rm min}=100$ due to a contaminant from residual foregrounds in the CIB map. We find that the SNR of the cross-correlation is $79$ with $f_{\rm sky}^{\phi I}=0.60$, which is roughly a factor of two improvement compared with obtained by \Planck.

\subsubsection{ISW-lensing cross-correlation}

The ISW effect provides information on large-scale structure through large-scale temperature fluctuations \cite{Sachs:1967}. As the same structure generates both the CMB lensing potential and the ISW effect, a substantial cross-correlation between the two observables is expected \cite{Goldberg:1999xm}. The cross-correlation between ISW and CMB lensing is only significant at low multipoles (see, e.g., Ref.~\cite{Lewis:2006:review}). Thus, to measure the cross-correlation with the ISW effect, we need a nearly full-sky observation of the lensing map from space, such as \Planck\ and \LB. We compute the SNR of the ISW-lensing cross-correlation signal, $C_{\ell}^{\phi T}$, assuming cosmic-variance limited temperature fluctuations and $f_{\rm sky}^{\phi T} = 0.80$. The estimated SNR for our baseline case is approximately $4$. 

\subsection{Constraints on tensor-to-scalar ratio by delensing}

Finally, we can use the internally reconstructed lensing potential for delensing. However, the lensing map of \LB\ does not significantly improve the constraint on this tensor-to-scalar ratio as shown in \paperII. The improvement on this constraint is only at the level of a few percent if we use the \LB\ lensing map (see \paperII\ for the details). 

\section{Summary and Discussion} \label{sec:summary}

We have conducted a lensing reconstruction study assuming the \LB\ experimental configuration, focusing on the impact of Galactic foregrounds from synchrotron and dust emission on the lensing analysis. We performed component separation on the frequency maps, applied filtering to the post-component-separated maps, and estimated the lensing potential map. We showed that the foreground-induced mean field and trispectrum have negligible biases on the measurement of the lensing power spectrum. Furthermore, we showed that the SNR of the \LB\ lensing map is approximately 40 which is comparable to that obtained from the latest \Planck\ measurement. The \LB\ lensing map additionally holds potential for several cross-correlation analyses.

We focused on the lensing reconstruction from \LB\ data alone, but we can add \Planck\ data to provide a more precise lensing map over the full sky, which will be investigated in future work. We assumed homogeneous white noise, but the \LB\ noise is inhomogeneous due to the scan strategy. This effect could introduce a larger mean-field bias \cite{Hanson:2009:noise,P18:phi}. However, the $EB$ estimator has no significant mean-field bias due to the parity symmetry \cite{namikawa-MF,Pearson:2014}. The inhomogeneous noise makes the reconstruction sub-optimal without including its effect in the analysis. However, we can optimally perform component-separation and lensing analysis by modifying the covariance matrix, including the inhomogeneity of the noise \cite{P18:phi}. For the optimal lensing analysis, we can also incorporate the spatial variation in the estimator normalization \cite{Mirmelstein:2019:filter}. 
We have also ignored the instrumental systematic effect of the \LB\ experiment. Beam systematics could be one of the main sources of biases in the lensing measurement since the lensing analysis uses smaller scales available for a given dataset. These practical issues, including an optimal analysis for inhomogeneous noise and residual foregrounds, and instrumental systematics, will be investigated in our future works. 
The software used for this study is publicly available at \url{https://github.com/litebird/LiteBIRD-lensing}.

\section*{Acknowledgments}
%
This work is supported in Japan by ISAS/JAXA for Pre-Phase A2 studies, by the acceleration program of JAXA research and development directorate, by the World Premier International Research Center Initiative (WPI) of MEXT, by the JSPS Core-to-Core Program of A. Advanced Research Networks, and by JSPS KAKENHI Grant Numbers JP15H05891, JP17H01115, and JP17H01125.
The Canadian contribution is supported by the Canadian Space Agency.
The French \textit{LiteBIRD} phase A contribution is supported by the Centre National d’Etudes Spatiale (CNES), by the Centre National de la Recherche Scientifique (CNRS), and by the Commissariat à l’Energie Atomique (CEA).
The German participation in \textit{LiteBIRD} is supported in part by the Excellence Cluster ORIGINS, which is funded by the Deutsche Forschungsgemeinschaft (DFG, German Research Foundation) under Germany’s Excellence Strategy (Grant No. EXC-2094 - 390783311).
The Italian \textit{LiteBIRD} phase A contribution is supported by the Italian Space Agency (ASI Grants No. 2020-9-HH.0 and 2016-24-H.1-2018), the National Institute for Nuclear Physics (INFN) and the National Institute for Astrophysics (INAF).
Norwegian participation in \textit{LiteBIRD} is supported by the Research Council of Norway (Grant No. 263011) and has received funding from the European Research Council (ERC) under the Horizon 2020 Research and Innovation Programme (Grant agreement No. 772253 and 819478).
The Spanish \textit{LiteBIRD} phase A contribution is supported by the Spanish Agencia Estatal de Investigación (AEI), project refs. PID2019-110610RB-C21,  PID2020-120514GB-I00, ProID2020010108 and ICTP20210008.
Funds that support contributions from Sweden come from the Swedish National Space Agency (SNSA/Rymdstyrelsen) and the Swedish Research Council (Reg. no. 2019-03959).
The US contribution is supported by NASA grant no. 80NSSC18K0132.
%
We also acknowledge the support from H2020-MSCA-RISE- 2020 European grant (Marie Sklodowska-Curie Research and Innovation Staff Exchange), JSPS KAKENHI Grant No. JP20H05859 and No. JP22K03682. 
This research used resources of the National Energy Research Scientific Computing Center (NERSC), a U.S. Department of Energy Office of Science User Facility located at Lawrence Berkeley National Laboratory.

\appendix

\section{Constraint on \texorpdfstring{$A_\mathrm{lens}$}{Alens}} 
\label{sec:Appendix}

\begin{figure}[ht]
 \centering
 \includegraphics[width=.5\textwidth]{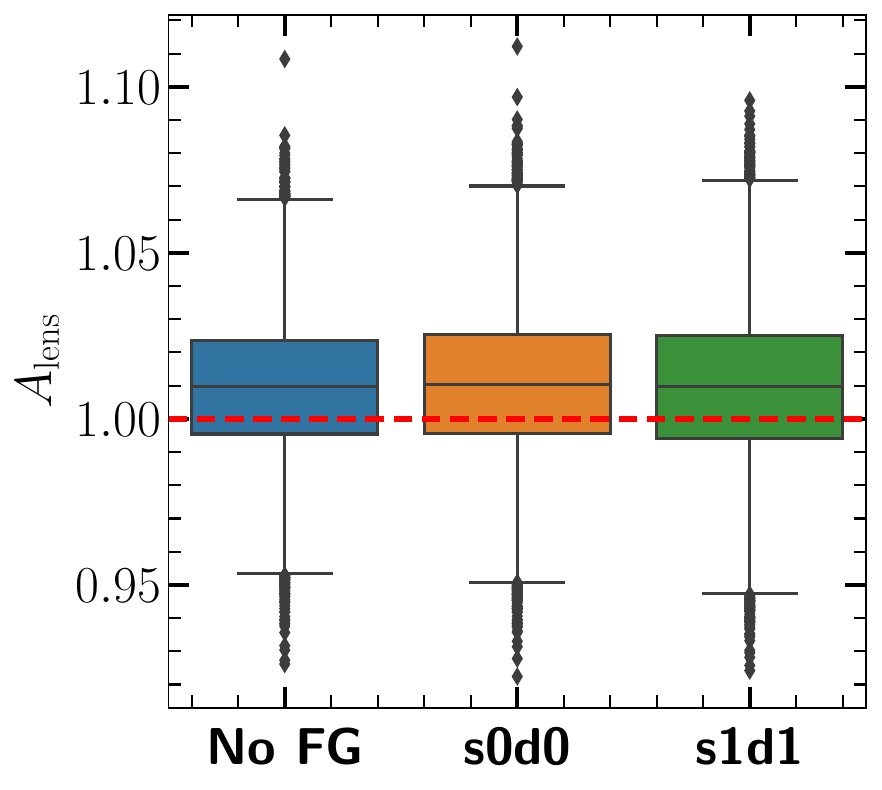}
 \caption{
 Box plots showing the distribution of samples, with blue, orange, and green denoting the no-foreground, \textbf{s0d0}, and \textbf{s1d1} cases, respectively. A red dashed line serves as the reference value, indicating $A_\mathrm{lens} = 1$. Within each box plot, a box spans from the first quartile to the third quartile, while a horizontal black line extends through the box at the median. Lower and upper whiskers correspond to $4\,\sigma$, and individual points denote outliers ($\geq 4\,\sigma$).
 }
 \label{fig:alensbox}
\end{figure}

This appendix shows some details on estimating $A_{\rm lens}$. We first note that Eq.~\eqref{Eq:AL} is motivated by the following likelihood:
\begin{equation}
    -2\ln(\mathcal{L}) = \sum_{bb'}
    \left(\hat{C}_b^{\phi\phi} - A_{\rm lens}C_b^{\phi\phi,{\rm theory}}\right) \{\mathbf{Cov}^{-1}\}_{bb'} \left(\hat{C}_{b'}^{\phi\phi} - A_{\rm lens} C_{b'}^{\phi\phi,{\rm theory}}\right)
    \,. 
\end{equation}
Differentiating the above likelihood in terms of $A_{\rm lens}$ leads to Eq.~\eqref{Eq:AL}. 
Instead of using Eq.~\eqref{Eq:AL}, we can also estimate $A_{\rm lens}$ by maximizing the above likelihood. 
For example, Fig.~\ref{fig:alensbox} shows the results with an Affine-Invariant Markov Chain Monte Carlo (MCMC) Ensemble sampler implemented in Python package $\textsc{emcee}$\footnote{\url{https://github.com/dfm/emcee}} for three different foreground cases. Note that the estimated $A_{\rm lens}$ values are in agreement with $A_\mathrm{lens}=1$ within the statistical errors. This highlights the agreement between our reconstructed lensing power spectrum in Eq.~\eqref{eq:cl_pp_final} and the theoretical input power spectrum.

\begin{figure}[ht!]
    \centering
    \includegraphics[width=0.8\textwidth]{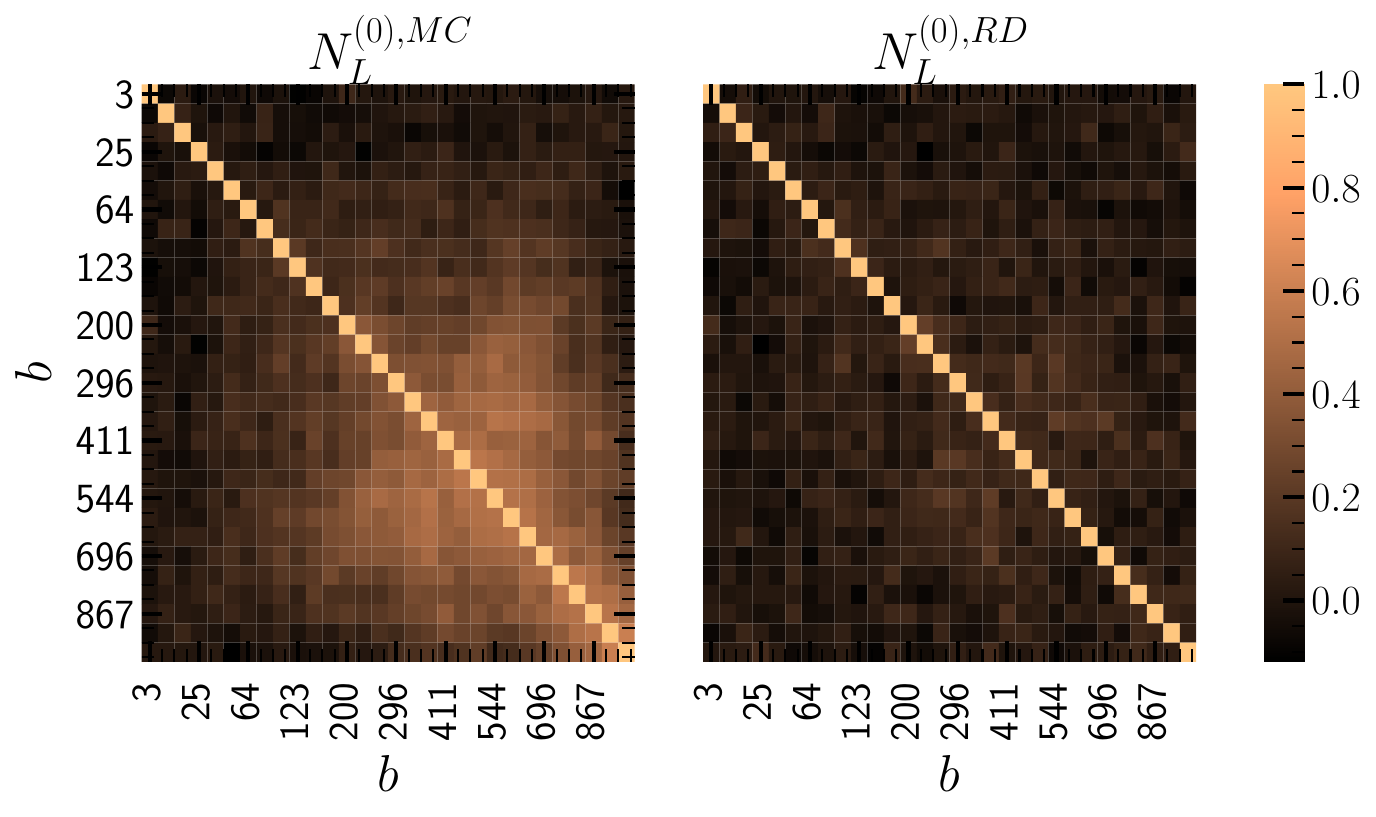}
    \caption{
    Correlation matrix of the binned angular power spectra for the \textbf{s1d1} case. The left panel represents results obtained using MCN0, while the right panel illustrates outcomes with RDN0. Notably, the utilization of RDN0 leads to a reduction in correlations within the angular power spectra.
    }
    \label{fig:corr_coef}
\end{figure}

To create the covariance matrix $\mathbf{Cov}$, one can use MCN0 into Eq.~\eqref{eq:cl_pp_final} as the estimate of the disconnected bias. However, this choice introduces correlations within the covariance matrix, which, in turn, influences the precision of the SNR estimation (e.g., Refs.~\cite{Schmittfull:2013,BKVIII,Peloton:2016:RDN0}). 
Instead, RDN0 has been used to reduce these correlations. For the \LB\ case, this reduction becomes visually evident when examining the correlation matrix shown in Fig.~\ref{fig:corr_coef}.

\bibliographystyle{Class/JHEP}
\bibliography{Misc/cite}

\providecommand{\href}[2]{#2}\begingroup\raggedright\begin{thebibliography}{100}

\bibitem{Zaldarriaga:1996:EBdef}
M.~Zaldarriaga and U.~Seljak, \emph{{An all sky analysis of polarization in the
  microwave background}},
  \href{https://doi.org/10.1103/PhysRevD.55.1830}{\emph{\prd} {\bfseries D55}
  (1997) 1830} [\href{https://arxiv.org/abs/astro-ph/9609170}{{\ttfamily
  astro-ph/9609170}}].

\bibitem{Kamionkowski_1997}
M.~Kamionkowski, A.~Kosowsky and A.~Stebbins, \emph{Statistics of cosmic
  microwave background polarization},
  \href{https://doi.org/10.1103/physrevd.55.7368}{\emph{\prd} {\bfseries 55}
  (1997) 7368}.

\bibitem{Seljak:1996ti}
U.~Seljak, \emph{{Measuring polarization in cosmic microwave background}},
  \href{https://doi.org/10.1086/304123}{\emph{\apj} {\bfseries 482} (1997) 6}
  [\href{https://arxiv.org/abs/astro-ph/9608131}{{\ttfamily
  astro-ph/9608131}}].

\bibitem{Seljak:1996gy}
U.~Seljak and M.~Zaldarriaga, \emph{Signature of gravity waves in polarization
  of the microwave background},
  \href{https://doi.org/10.1103/PhysRevLett.78.2054}{\emph{\prl} {\bfseries 78}
  (1997) 2054} [\href{https://arxiv.org/abs/astro-ph/9609169}{{\ttfamily
  astro-ph/9609169}}].

\bibitem{Kamionkowski:1996zd}
M.~Kamionkowski, A.~Kosowsky and A.~Stebbins, \emph{{A Probe of primordial
  gravity waves and vorticity}},
  \href{https://doi.org/10.1103/PhysRevLett.78.2058}{\emph{\prl} {\bfseries 78}
  (1997) 2058} [\href{https://arxiv.org/abs/astro-ph/9609132}{{\ttfamily
  astro-ph/9609132}}].

\bibitem{LiteBIRD}
{\scshape LiteBIRD} collaboration, \emph{{Probing Cosmic Inflation with the
  LiteBIRD Cosmic Microwave Background Polarization Survey}},
  \href{https://doi.org/10.1093/ptep/ptac150}{\emph{\ptep} {\bfseries 2023}
  (2023) 042F01} [\href{https://arxiv.org/abs/2202.02773}{{\ttfamily
  2202.02773}}].

\bibitem{Zaldarriaga:1998:LensB}
M.~Zaldarriaga and U.~Seljak, \emph{{Gravitational lensing effect on cosmic
  microwave background polarization}},
  \href{https://doi.org/10.1103/PhysRevD.58.023003}{\emph{\prd} {\bfseries 58}
  (1998) 023003} [\href{https://arxiv.org/abs/astro-ph/9803150}{{\ttfamily
  astro-ph/9803150}}].

\bibitem{Smith:2007}
K.M.~Smith, O.~Zahn and O.~Dore, \emph{Detection of gravitational lensing in
  the cosmic microwave background},
  \href{https://doi.org/10.1103/PhysRevD.76.043510}{\emph{\prd} {\bfseries 76}
  (2007) 043510} [\href{https://arxiv.org/abs/0705.3980}{{\ttfamily
  0705.3980}}].

\bibitem{Hirata:2008}
C.M.~Hirata, S.~Ho, N.~Padmanabhan, U.~Seljak and N.A.~Bahcall,
  \emph{{Correlation of CMB with large-scale structure: II. Weak lensing}},
  \href{https://doi.org/10.1103/PhysRevD.78.043520}{\emph{\prd} {\bfseries 78}
  (2008) 043520} [\href{https://arxiv.org/abs/0801.0644}{{\ttfamily
  0801.0644}}].

\bibitem{Bianchini:2014}
F.~Bianchini et~al., \emph{{Cross-correlation between the CMB lensing potential
  measured by Planck and high-z sub-mm galaxies detected by the Herschel-ATLAS
  survey}}, \href{https://doi.org/10.1088/0004-637X/802/1/64}{\emph{\apj}
  {\bfseries 802} (2015) 64} [\href{https://arxiv.org/abs/1410.4502}{{\ttfamily
  1410.4502}}].

\bibitem{Giannantonio:DES:2015}
{\scshape DES} collaboration, \emph{{CMB lensing tomography with the DES
  Science Verification galaxies}},
  \href{https://doi.org/10.1093/mnras/stv2678}{\emph{\mnras} {\bfseries 456}
  (2016) 3213} [\href{https://arxiv.org/abs/1507.05551}{{\ttfamily
  1507.05551}}].

\bibitem{Singh:2016xey}
S.~Singh, R.~Mandelbaum and J.R.~Brownstein, \emph{{Cross-correlating Planck
  CMB lensing with SDSS: Lensing-lensing and galaxy-lensing
  cross-correlations}},
  \href{https://doi.org/10.1093/mnras/stw2482}{\emph{\mnras} {\bfseries 464}
  (2017) 2120} [\href{https://arxiv.org/abs/1606.08841}{{\ttfamily
  1606.08841}}].

\bibitem{Omori:DES:2018:gal}
{\scshape DES, SPT} collaboration, \emph{{Dark Energy Survey Year 1 Results:
  Tomographic cross-correlations between Dark Energy Survey galaxies and CMB
  lensing from South Pole Telescope+Planck}},
  \href{https://doi.org/10.1103/PhysRevD.100.043501}{\emph{\prd} {\bfseries
  100} (2019) 043501} [\href{https://arxiv.org/abs/1810.02342}{{\ttfamily
  1810.02342}}].

\bibitem{Polarbear:2019:gal}
{\scshape POLARBEAR} collaboration, \emph{{Cross-correlation of POLARBEAR CMB
  Polarization Lensing with High-$z$ Sub-mm Herschel-ATLAS galaxies}},
  \href{https://doi.org/10.3847/1538-4357/ab4a78}{\emph{\apj} {\bfseries 886}
  (2019) 38} [\href{https://arxiv.org/abs/1903.07046}{{\ttfamily 1903.07046}}].

\bibitem{Marques:2019}
G.A.~Marques and A.~Bernui, \emph{{Tomographic analyses of the CMB lensing and
  galaxy clustering to probe the linear structure growth}},
  \href{https://doi.org/10.1088/1475-7516/2020/05/052}{\emph{\jcap} {\bfseries
  05} (2020) 052} [\href{https://arxiv.org/abs/1908.04854}{{\ttfamily
  1908.04854}}].

\bibitem{Darwish:2020}
O.~Darwish, M.S.~Madhavacheril, B.~Sherwin, S.~Aiola, N.~Battaglia, J.A.~Beall
  et~al., \emph{{The Atacama Cosmology Telescope: A CMB lensing mass map over
  2100 square degrees of sky and its cross-correlation with BOSS-CMASS
  galaxies}}, \href{https://doi.org/10.1093/mnras/staa3438}{\emph{\mnras}
  {\bfseries 500} (2020) 2250}
  [\href{https://arxiv.org/abs/2004.01139}{{\ttfamily 2004.01139}}].

\bibitem{Krolewski:2021}
A.~Krolewski, S.~Ferraro and M.~White, \emph{{Cosmological constraints from
  unWISE and Planck CMB lensing tomography}},
  \href{https://doi.org/10.1088/1475-7516/2021/12/028}{\emph{\jcap} {\bfseries
  12} (2021) 028} [\href{https://arxiv.org/abs/2105.03421}{{\ttfamily
  2105.03421}}].

\bibitem{Dong:2021:void}
F.~Dong, P.~Zhang, L.~Zhang, J.~Yao, Z.~Sun, C.~Park et~al., \emph{{Detection
  of a Cross-correlation between Cosmic Microwave Background Lensing and
  Low-density Points}},
  \href{https://doi.org/10.3847/1538-4357/ac2d31}{\emph{\apj} {\bfseries 923}
  (2021) 153} [\href{https://arxiv.org/abs/2107.08694}{{\ttfamily
  2107.08694}}].

\bibitem{Sun:2021}
Z.~Sun, J.~Yao, F.~Dong, X.~Yang, L.~Zhang and P.~Zhang,
  \emph{{Cross-correlation of Planck cosmic microwave background lensing with
  DESI galaxy groups}},
  \href{https://doi.org/10.1093/mnras/stac138}{\emph{\mnras} {\bfseries 511}
  (2022) 3548} [\href{https://arxiv.org/abs/2109.07387}{{\ttfamily
  2109.07387}}].

\bibitem{Miyatake:2021}
H.~Miyatake, Y.~Harikane, M.~Ouchi, Y.~Ono, N.~Yamamoto, A.J.~Nishizawa et~al.,
  \emph{{First Identification of a CMB Lensing Signal Produced by 1.5~Million
  Galaxies at z\ensuremath{\sim}4: Constraints on Matter Density Fluctuations
  at High Redshift}},
  \href{https://doi.org/10.1103/PhysRevLett.129.061301}{\emph{\prl} {\bfseries
  129} (2022) 061301} [\href{https://arxiv.org/abs/2103.15862}{{\ttfamily
  2103.15862}}].

\bibitem{Saraf:2021}
C.S.~Saraf, P.~Bielewicz and M.~Chodorowski, \emph{{Cross-correlation between
  Planck CMB lensing potential and galaxy catalogues from HELP}},
  \href{https://doi.org/10.1093/mnras/stac1876}{\emph{\mnras} {\bfseries 515}
  (2022) 1993} [\href{https://arxiv.org/abs/2106.02551}{{\ttfamily
  2106.02551}}].

\bibitem{Omori:DES:2022}
{\scshape DES, SPT} collaboration, \emph{{Joint analysis of Dark Energy Survey
  Year 3 data and CMB lensing from SPT and Planck. I. Construction of CMB
  lensing maps and modeling choices}},
  \href{https://doi.org/10.1103/PhysRevD.107.023529}{\emph{\prd} {\bfseries
  107} (2023) 023529} [\href{https://arxiv.org/abs/2203.12439}{{\ttfamily
  2203.12439}}].

\bibitem{Chang:DES:2022}
{\scshape DES, SPT} collaboration, \emph{{Joint analysis of Dark Energy Survey
  Year 3 data and CMB lensing from SPT and Planck. II. Cross-correlation
  measurements and cosmological constraints}},
  \href{https://doi.org/10.1103/PhysRevD.107.023530}{\emph{\prd} {\bfseries
  107} (2023) 023530} [\href{https://arxiv.org/abs/2203.12440}{{\ttfamily
  2203.12440}}].

\bibitem{Piccirilli:2022:phixgal}
G.~Piccirilli, M.~Migliaccio, E.~Branchini and A.~Dolfi, \emph{{A
  cross-correlation analysis of CMB lensing and radio galaxy maps}},
  \href{https://doi.org/10.1051/0004-6361/202244799}{\emph{\aap} {\bfseries
  671} (2023) A42} [\href{https://arxiv.org/abs/2208.07774}{{\ttfamily
  2208.07774}}].

\bibitem{Yao:2023}
J.~Yao et~al., \emph{{KiDS-1000: cross-correlation with Planck cosmic microwave
  background lensing and intrinsic alignment removal with self-calibration}},
  \href{https://doi.org/10.1051/0004-6361/202346020}{\emph{\aap} {\bfseries
  673} (2023) A111} [\href{https://arxiv.org/abs/2301.13437}{{\ttfamily
  2301.13437}}].

\bibitem{Farren:ACT:2023}
{\scshape ACT} collaboration, \emph{{The Atacama Cosmology Telescope: Cosmology
  from cross-correlations of unWISE galaxies and ACT DR6 CMB lensing}},
  {\emph{{}} (2023) } [\href{https://arxiv.org/abs/2309.05659}{{\ttfamily
  2309.05659}}].

\bibitem{Farren:2023}
G.S.~Farren, B.D.~Sherwin, B.~Bolliet, T.~Namikawa, S.~Ferraro and
  A.~Krolewski, \emph{{Detection of the CMB lensing -- galaxy bispectrum}},
  {\emph{{}} (2023) } [\href{https://arxiv.org/abs/2311.04213}{{\ttfamily
  2311.04213}}].

\bibitem{Hand:2013:phixshear}
N.~Hand, A.~Leauthaud, S.~Das, B.D.~Sherwin, G.E.~Addison, J.R.~Bond et~al.,
  \emph{{First measurement of the cross-correlation of CMB lensing and galaxy
  lensing}}, \href{https://doi.org/10.1103/PhysRevD.91.062001}{\emph{\prd}
  {\bfseries 91} (2015) 062001}
  [\href{https://arxiv.org/abs/1311.6200}{{\ttfamily 1311.6200}}].

\bibitem{Kirk:DES:2015}
{\scshape DES} collaboration, \emph{{Cross-correlation of gravitational lensing
  from DES Science Verification data with SPT and Planck lensing}},
  \href{https://doi.org/10.1093/mnras/stw570}{\emph{\mnras} {\bfseries 459}
  (2016) 21} [\href{https://arxiv.org/abs/1512.04535}{{\ttfamily 1512.04535}}].

\bibitem{Liu:2015}
J.~Liu and J.C.~Hill, \emph{{Cross-correlation of Planck CMB Lensing and
  CFHTLenS Galaxy Weak Lensing Maps}},
  \href{https://doi.org/10.1103/PhysRevD.92.063517}{\emph{\prd} {\bfseries 92}
  (2015) 063517} [\href{https://arxiv.org/abs/1504.05598}{{\ttfamily
  1504.05598}}].

\bibitem{Harnois-Deraps:2016}
J.~Harnois-D\'eraps et~al., \emph{{CFHTLenS and RCSLenS Cross-Correlation with
  Planck Lensing Detected in Fourier and Configuration Space}},
  \href{https://doi.org/10.1093/mnras/stw947}{\emph{\mnras} {\bfseries 460}
  (2016) 434} [\href{https://arxiv.org/abs/1603.07723}{{\ttfamily
  1603.07723}}].

\bibitem{Harnois-Deraps:2017kfd}
J.~Harnois-D\'eraps et~al., \emph{{KiDS-450: tomographic cross-correlation of
  galaxy shear with Planck lensing}},
  \href{https://doi.org/10.1093/mnras/stx1675}{\emph{\mnras} {\bfseries 471}
  (2017) 1619} [\href{https://arxiv.org/abs/1703.03383}{{\ttfamily
  1703.03383}}].

\bibitem{Omori:DES:2018:shear}
{\scshape DES, SPT} collaboration, \emph{{Dark Energy Survey Year 1 Results:
  Cross-correlation between Dark Energy Survey Y1 galaxy weak lensing and South
  Pole Telescope +Planck CMB weak lensing}},
  \href{https://doi.org/10.1103/PhysRevD.100.043517}{\emph{\prd} {\bfseries
  100} (2019) 043517} [\href{https://arxiv.org/abs/1810.02441}{{\ttfamily
  1810.02441}}].

\bibitem{Namikawa:2019:hscxpb}
{\scshape POLARBEAR, HSC} collaboration, \emph{{Evidence for the
  Cross-correlation between Cosmic Microwave Background Polarization Lensing
  from POLARBEAR and Cosmic Shear from Subaru Hyper Suprime-Cam}},
  \href{https://doi.org/10.3847/1538-4357/ab3424}{\emph{\apj} {\bfseries 882}
  (2019) 62} [\href{https://arxiv.org/abs/1904.02116}{{\ttfamily 1904.02116}}].

\bibitem{ACT:Robertson:2020}
N.C.~Robertson, D.~Alonso, J.~{Harnois-D{\'e}raps}, O.~Darwish, A.~Kannawadi,
  A.~Amon et~al., \emph{{Strong detection of the CMB lensing and galaxy weak
  lensing cross-correlation from ACT-DR4, Planck Legacy, and KiDS-1000}},
  \href{https://doi.org/10.1051/0004-6361/202039975}{\emph{\aap} {\bfseries
  649} (2021) A146} [\href{https://arxiv.org/abs/2011.11613}{{\ttfamily
  2011.11613}}].

\bibitem{Marques:2020}
G.A.~Marques, J.~Liu, K.M.~Huffenberger and J.~Colin~Hill,
  \emph{{Cross-correlation between Subaru Hyper Suprime-Cam Galaxy Weak Lensing
  and Planck Cosmic Microwave Background Lensing}},
  \href{https://doi.org/10.3847/1538-4357/abc003}{\emph{\apj} {\bfseries 904}
  (2020) 182} [\href{https://arxiv.org/abs/2008.04369}{{\ttfamily
  2008.04369}}].

\bibitem{Shaikh:ACT:2023}
{\scshape ACT, DES} collaboration, \emph{{Cosmology from Cross-Correlation of
  ACT-DR4 CMB Lensing and DES-Y3 Cosmic Shear}}, {\emph{{}} (2023) }
  [\href{https://arxiv.org/abs/2309.04412}{{\ttfamily 2309.04412}}].

\bibitem{P13:ISW}
{\textit{Planck} Collaboration}, \emph{{Planck 2013 results. XIX. The
  integrated Sachs-Wolfe effect}},
  \href{https://doi.org/10.1051/0004-6361/201321526}{\emph{\aap} {\bfseries
  571} (2014) A19} [\href{https://arxiv.org/abs/1303.5079}{{\ttfamily
  1303.5079}}].

\bibitem{P15:ISW}
{\textit{Planck} Collaboration}, \emph{{Planck 2015 results. XXI. The
  integrated Sachs-Wolfe effect}},
  \href{https://doi.org/10.1051/0004-6361/201525831}{\emph{\aap} {\bfseries
  594} (2016) A21} [\href{https://arxiv.org/abs/1502.01595}{{\ttfamily
  1502.01595}}].

\bibitem{P18:phi}
{\textit{Planck} Collaboration}, \emph{{Planck 2018 results. VIII.
  Gravitational lensing}},
  \href{https://doi.org/10.1051/0004-6361/201833886}{\emph{\aap} {\bfseries
  641} (2020) A8} [\href{https://arxiv.org/abs/1807.06210}{{\ttfamily
  1807.06210}}].

\bibitem{Carron:2022:NPIPE-phixISW}
J.~Carron, A.~Lewis and G.~Fabbian, \emph{{Planck integrated
  Sachs-Wolfe-lensing likelihood and the CMB temperature}},
  \href{https://doi.org/10.1103/PhysRevD.106.103507}{\emph{\prd} {\bfseries
  106} (2022) 103507} [\href{https://arxiv.org/abs/2209.07395}{{\ttfamily
  2209.07395}}].

\bibitem{Hill:2013}
J.C.~Hill and D.N.~Spergel, \emph{{Detection of thermal SZ-CMB lensing
  cross-correlation in Planck nominal mission data}},
  \href{https://doi.org/10.1088/1475-7516/2014/02/030}{\emph{\jcap} {\bfseries
  02} (2014) 030} [\href{https://arxiv.org/abs/1312.4525}{{\ttfamily
  1312.4525}}].

\bibitem{McCarthy:2023:tSZ}
F.~McCarthy and J.C.~Hill, \emph{{Cross-correlation of the thermal
  Sunyaev--Zel'dovich and CMB lensing signals in Planck PR4 data with robust
  CIB decontamination}}, {\emph{{}} (2023) }
  [\href{https://arxiv.org/abs/2308.16260}{{\ttfamily 2308.16260}}].

\bibitem{Holder:2013:phixCIB}
G.P.~{Holder}, M.P.~{Viero}, O.~{Zahn}, K.A.~{Aird}, B.A.~{Benson},
  S.~{Bhattacharya} et~al., \emph{{A Cosmic Microwave Background Lensing Mass
  Map and Its Correlation with the Cosmic Infrared Background}},
  \href{https://doi.org/10.1088/2041-8205/771/1/L16}{\emph{\apj} {\bfseries
  771} (2013) L16} [\href{https://arxiv.org/abs/1303.5048}{{\ttfamily
  1303.5048}}].

\bibitem{P13:CIBxphi}
{\textit{Planck} Collaboration}, \emph{{Planck 2013 results. XVIII. The
  gravitational lensing-infrared background correlation}},
  \href{https://doi.org/10.1051/0004-6361/201321540}{\emph{\aap} {\bfseries
  571} (2014) A18} [\href{https://arxiv.org/abs/1303.5078}{{\ttfamily
  1303.5078}}].

\bibitem{Hanson:2013:lensB}
D.~Hanson, S.~Hoover, A.~Crites, P.A.R.~Ade, K.A.~Aird, J.E.~Austermann et~al.,
  \emph{{Detection of B-mode Polarization in the Cosmic Microwave Background
  with Data from the South Pole Telescope}},
  \href{https://doi.org/10.1103/PhysRevLett.111.141301}{\emph{\prl} {\bfseries
  111} (2013) 141301} [\href{https://arxiv.org/abs/1307.5830}{{\ttfamily
  1307.5830}}].

\bibitem{PB:2014:phixI}
{\scshape POLARBEAR} collaboration, \emph{{Evidence for Gravitational Lensing
  of the Cosmic Microwave Background Polarization from Cross-Correlation with
  the Cosmic Infrared Background}},
  \href{https://doi.org/10.1103/PhysRevLett.112.131302}{\emph{\prl} {\bfseries
  112} (2014) 131302} [\href{https://arxiv.org/abs/1312.6645}{{\ttfamily
  1312.6645}}].

\bibitem{vanEngelen:ACT:2014:cib}
{\scshape ACT} collaboration, \emph{{The Atacama Cosmology Telescope: Lensing
  of CMB Temperature and Polarization Derived from Cosmic Infrared Background
  Cross-Correlation}},
  \href{https://doi.org/10.1088/0004-637X/808/1/7}{\emph{\apj} {\bfseries 808}
  (2015) 7} [\href{https://arxiv.org/abs/1412.0626}{{\ttfamily 1412.0626}}].

\bibitem{Cao:2019:phixI}
Y.~Cao, Y.~Gong, C.~Feng, A.~Cooray, G.~Cheng and X.~Chen,
  \emph{{Cross-Correlation of Far-Infrared Background Anisotropies and CMB
  Lensing from Herschel and Planck satellites}},
  \href{https://doi.org/10.3847/1538-4357/abada1}{\emph{\apj} {\bfseries 901}
  (2020) 34} [\href{https://arxiv.org/abs/1912.12840}{{\ttfamily 1912.12840}}].

\bibitem{Kesden:2002:delens}
M.H.~Kesden, A.~Cooray and M.~Kamionkowski, \emph{{Separation of gravitational
  wave and cosmic shear contributions to cosmic microwave background
  polarization}},
  \href{https://doi.org/10.1103/PhysRevLett.89.011304}{\emph{\prl} {\bfseries
  89} (2002) 011304} [\href{https://arxiv.org/abs/astro-ph/0202434}{{\ttfamily
  astro-ph/0202434}}].

\bibitem{Seljak:2003pn}
U.~Seljak and C.M.~Hirata, \emph{Gravitational lensing as a contaminant of the
  gravity wave signal in cmb},
  \href{https://doi.org/10.1103/PhysRevD.69.043005}{\emph{\prd} {\bfseries 69}
  (2004) 043005} [\href{https://arxiv.org/abs/astro-ph/0310163}{{\ttfamily
  astro-ph/0310163}}].

\bibitem{Smith:2010:delens}
K.M.~Smith, D.~Hanson, M.~LoVerde, C.M.~Hirata and O.~Zahn, \emph{{Delensing
  CMB Polarization with External Datasets}},
  \href{https://doi.org/10.1088/1475-7516/2012/06/014}{\emph{\jcap} {\bfseries
  06} (2012) 014} [\href{https://arxiv.org/abs/1010.0048}{{\ttfamily
  1010.0048}}].

\bibitem{OkamotoHu:quad}
T.~Okamoto and W.~Hu, \emph{{CMB Lensing Reconstruction on the Full Sky}},
  \href{https://doi.org/10.1103/PhysRevD.67.083002}{\emph{\prd} {\bfseries 67}
  (2003) 083002} [\href{https://arxiv.org/abs/astro-ph/0301031}{{\ttfamily
  astro-ph/0301031}}].

\bibitem{HuOkamoto:2001}
W.~Hu and T.~Okamoto, \emph{Mass reconstruction with cmb polarization},
  \href{https://doi.org/10.1086/341110}{\emph{\apj} {\bfseries 574} (2002) 566}
  [\href{https://arxiv.org/abs/astro-ph/0111606}{{\ttfamily
  astro-ph/0111606}}].

\bibitem{Das:2011}
S.~Das, B.D.~Sherwin, P.~Aguirre, J.W.~Appel, J.R.~Bond, C.S.~Carvalho et~al.,
  \emph{{Detection of the Power Spectrum of Cosmic Microwave Background Lensing
  by the Atacama Cosmology Telescope}},
  \href{https://doi.org/10.1103/PhysRevLett.107.021301}{\emph{\prl} {\bfseries
  107} (2011) 021301} [\href{https://arxiv.org/abs/1103.2124}{{\ttfamily
  1103.2124}}].

\bibitem{Sherwin:2011:DE}
B.D.~Sherwin et~al., \emph{Evidence for dark energy from the cosmic microwave
  background alone using the atacama cosmology telescope lensing measurements},
  {\emph{\prl} {\bfseries 107} (2011) 021302}
  [\href{https://arxiv.org/abs/1105.0419}{{\ttfamily 1105.0419}}].

\bibitem{ACT16:phi}
B.D.~{Sherwin}, A.~{van Engelen}, N.~{Sehgal}, M.~{Madhavacheril},
  G.E.~{Addison}, S.~{Aiola} et~al., \emph{{The Atacama Cosmology Telescope:
  Two-Season ACTPol Lensing Power Spectrum}},
  \href{https://doi.org/10.1103/PhysRevD.95.123529}{\emph{\prd} {\bfseries 95}
  (2017) 123529} [\href{https://arxiv.org/abs/1611.09753}{{\ttfamily
  1611.09753}}].

\bibitem{Qu:ACT:2023}
{\scshape ACT} collaboration, \emph{{The Atacama Cosmology Telescope: A
  Measurement of the DR6 CMB Lensing Power Spectrum and its Implications for
  Structure Growth}},  \href{https://arxiv.org/abs/2304.05202}{{\ttfamily
  2304.05202}}.

\bibitem{Madhavacheril:ACT:2023}
{\scshape ACT} collaboration, \emph{{The Atacama Cosmology Telescope: DR6
  Gravitational Lensing Map and Cosmological Parameters}},
  \href{https://arxiv.org/abs/2304.05203}{{\ttfamily 2304.05203}}.

\bibitem{MacCrann:ACT:2023}
{\scshape ACT} collaboration, \emph{{The Atacama Cosmology Telescope:
  Mitigating the impact of extragalactic foregrounds for the DR6 CMB lensing
  analysis}},  \href{https://arxiv.org/abs/2304.05196}{{\ttfamily 2304.05196}}.

\bibitem{BKVIII}
{\textsc{Bicep2} / {\it Keck Array} Collaboration}, \emph{{{\sc BICEP2} / {\it
  Keck Array} VIII: Measurement of gravitational lensing from large-scale
  B-mode polarization}},
  \href{https://doi.org/10.3847/1538-4357/833/2/228}{\emph{\apj} {\bfseries
  833} (2016) 228} [\href{https://arxiv.org/abs/1606.01968}{{\ttfamily
  1606.01968}}].

\bibitem{BICEPKeck:2022:los-dist}
{\textsc{Bicep2} / {\it Keck Array} Collaboration}, \emph{{BICEP / Keck XVII:
  Line of Sight Distortion Analysis: Estimates of Gravitational Lensing,
  Anisotropic Cosmic Birefringence, Patchy Reionization, and Systematic
  Errors}}, {\emph{{}} (2022) }
  [\href{https://arxiv.org/abs/2210.08038}{{\ttfamily 2210.08038}}].

\bibitem{P13:phi}
{\textit{Planck} Collaboration}, \emph{{Planck 2013 results. XVII.
  Gravitational lensing by large-scale structure}},
  \href{https://doi.org/10.1051/0004-6361/201321543}{\emph{\aap} {\bfseries
  571} (2014) A17} [\href{https://arxiv.org/abs/1303.5077}{{\ttfamily
  1303.5077}}].

\bibitem{P15:phi}
{\textit{Planck} Collaboration}, \emph{{Planck 2015 results. XV. Gravitational
  lensing}}, {\emph{\aap} {\bfseries 594} (2015) A15}
  [\href{https://arxiv.org/abs/1502.01591}{{\ttfamily 1502.01591}}].

\bibitem{Carron:2022:NPIPE-lensing}
J.~Carron, M.~Mirmelstein and A.~Lewis, \emph{{CMB lensing from Planck
  PR4~maps}}, \href{https://doi.org/10.1088/1475-7516/2022/09/039}{\emph{\jcap}
  {\bfseries 09} (2022) 039}
  [\href{https://arxiv.org/abs/2206.07773}{{\ttfamily 2206.07773}}].

\bibitem{PB:phi:2013}
{\scshape POLARBEAR} collaboration, \emph{{Measurement of the Cosmic Microwave
  Background Polarization Lensing Power Spectrum with the POLARBEAR
  experiment}},
  \href{https://doi.org/10.1103/PhysRevLett.113.021301}{\emph{\prl} {\bfseries
  113} (2014) 021301} [\href{https://arxiv.org/abs/1312.6646}{{\ttfamily
  1312.6646}}].

\bibitem{PB:phi:2019}
{\scshape POLARBEAR} collaboration, \emph{{Measurement of the Cosmic Microwave
  Background Polarization Lensing Power Spectrum from Two Years of POLARBEAR
  Data}}, \href{https://doi.org/10.3847/1538-4357/ab7e29}{\emph{\apj}
  {\bfseries 893} (2020) 85}
  [\href{https://arxiv.org/abs/1911.10980}{{\ttfamily 1911.10980}}].

\bibitem{SPT:phi:2012}
A.~{van Engelen}, R.~{Keisler}, O.~{Zahn}, K.A.~{Aird}, B.A.~{Benson},
  L.E.~{Bleem} et~al., \emph{{A Measurement of Gravitational Lensing of the
  Microwave Background Using South Pole Telescope Data}},
  \href{https://doi.org/10.1088/0004-637X/756/2/142}{\emph{\apj} {\bfseries
  756} (2012) 142} [\href{https://arxiv.org/abs/1202.0546}{{\ttfamily
  1202.0546}}].

\bibitem{Story:2015}
K.T.~Story, D.~Hanson, P.A.R.~Ade, K.A.~Aird, J.E.~Austermann, J.A.~Beall
  et~al., \emph{A measurement of the cosmic microwave background gravitational
  lensing potential from 100 square degrees of sptpol data},
  \href{https://doi.org/10.1088/0004-637X/810/1/50}{\emph{\apj} {\bfseries 810}
  (2015) 50}.

\bibitem{SPT:phi:2019}
W.L.K.~Wu et~al., \emph{{A Measurement of the Cosmic Microwave Background
  Lensing Potential and Power Spectrum from 500 deg$^{2}$ of SPTpol Temperature
  and Polarization Data}},
  \href{https://doi.org/10.3847/1538-4357/ab4186}{\emph{\apj} {\bfseries 884}
  (2019) 70} [\href{https://arxiv.org/abs/1905.05777}{{\ttfamily 1905.05777}}].

\bibitem{Millea:2020}
M.~Millea et~al., \emph{{Optimal Cosmic Microwave Background Lensing
  Reconstruction and Parameter Estimation with SPTpol Data}},
  \href{https://doi.org/10.3847/1538-4357/ac02bb}{\emph{Astrophys. J.}
  {\bfseries 922} (2021) 259}
  [\href{https://arxiv.org/abs/2012.01709}{{\ttfamily 2012.01709}}].

\bibitem{Pan:SPT:2023}
{\scshape SPT} collaboration, \emph{{A Measurement of Gravitational Lensing of
  the Cosmic Microwave Background Using SPT-3G 2018 Data}},
  \href{https://arxiv.org/abs/2308.11608}{{\ttfamily 2308.11608}}.

\bibitem{SimonsObservatory}
{Simons Observatory Collaboration}, \emph{{The Simons Observatory: Science
  goals and forecasts}},
  \href{https://doi.org/10.1088/1475-7516/2019/02/056}{\emph{\jcap} {\bfseries
  02} (2019) 056} [\href{https://arxiv.org/abs/1808.07445}{{\ttfamily
  1808.07445}}].

\bibitem{CMBS4}
{CMB-S4 Collaboration}, \emph{{CMB-S4 Science Case, Reference Design, and
  Project Plan}}, {\emph{arXiv:1907.04473} (2019) }
  [\href{https://arxiv.org/abs/1907.04473}{{\ttfamily 1907.04473}}].

\bibitem{AliCPT:phi}
J.~Liu, Z.~Sun, J.~Han, J.~Carron, J.~Delabrouille, S.~Li et~al.,
  \emph{{Forecasts on CMB lensing observations with AliCPT-1}},
  \href{https://doi.org/10.1007/s11433-022-1966-4}{\emph{Sci. China Phys. Mech.
  Astron.} {\bfseries 65} (2022) 109511}
  [\href{https://arxiv.org/abs/2204.08158}{{\ttfamily 2204.08158}}].

\bibitem{Perotto:2009}
L.~Perotto, J.~Bobin, S.~Plaszczynski, J.L.~Starck and A.~Lavabre,
  \emph{{Reconstruction of the CMB lensing for Planck}},
  \href{https://doi.org/10.48550/arXiv.0903.1308}{\emph{{}} (2009) }
  [\href{https://arxiv.org/abs/0903.1308}{{\ttfamily 0903.1308}}].

\bibitem{Plaszczynski:2012}
S.~Plaszczynski, A.~Lavabre, L.~Perotto and J.L.~Starck, \emph{{A hybrid
  approach to CMB lensing reconstruction on all-sky intensity maps}},
  \href{https://doi.org/10.1051/0004-6361/201218899}{\emph{\aap} {\bfseries
  544} (2012) A27} [\href{https://arxiv.org/abs/1201.5779}{{\ttfamily
  1201.5779}}].

\bibitem{Namikawa:2012:bhe}
T.~Namikawa, D.~Hanson and R.~Takahashi, \emph{{Bias-Hardened CMB Lensing}},
  \href{https://doi.org/10.1093/mnras/stt195}{\emph{\mnras} {\bfseries 431}
  (2013) 609} [\href{https://arxiv.org/abs/1209.0091}{{\ttfamily 1209.0091}}].

\bibitem{namikawa-MF}
T.~Namikawa and R.~Takahashi, \emph{{Bias-hardened CMB lensing with
  polarization}}, \href{https://doi.org/10.1093/mnras/stt2290}{\emph{Monthly
  Notices of the Royal Astronomical Society} {\bfseries 438} (2013) 1507}
  [\href{https://arxiv.org/abs/1310.2372}{{\ttfamily 1310.2372}}].

\bibitem{BenoitLevy:2013:cutsky}
A.~Benoit-Levy, T.~Dechelette, K.~Benabed, J.-F.~Cardoso, D.~Hanson and
  S.~Prunet, \emph{{Full-sky CMB lensing reconstruction in presence of
  sky-cuts}}, \href{https://doi.org/10.1051/0004-6361/201321048}{\emph{\aap}
  {\bfseries 555} (2013) A37}
  [\href{https://arxiv.org/abs/1301.4145}{{\ttfamily 1301.4145}}].

\bibitem{Hanson:2009:noise}
D.~Hanson, G.~Rocha and K.~Gorski, \emph{{Lensing reconstruction from PLANCK
  sky maps: inhomogeneous noise}}, {\emph{\mnras} {\bfseries 400} (2009) 2169}
  [\href{https://arxiv.org/abs/0907.1927}{{\ttfamily 0907.1927}}].

\bibitem{Osborne:2013nna}
S.J.~Osborne, D.~Hanson and O.~Dore, \emph{{Extragalactic Foreground
  Contamination in Temperature-based CMB Lens Reconstruction}},
  \href{https://doi.org/10.1088/1475-7516/2014/03/024}{\emph{\jcap} {\bfseries
  03} (2014) 024} [\href{https://arxiv.org/abs/1310.7547}{{\ttfamily
  1310.7547}}].

\bibitem{Schaan:2018}
E.~Schaan and S.~Ferraro, \emph{{Foreground-Immune Cosmic Microwave Background
  Lensing with Shear-Only Reconstruction}},
  \href{https://doi.org/10.1103/PhysRevLett.122.181301}{\emph{\prl} {\bfseries
  122} (2019) 181301} [\href{https://arxiv.org/abs/1804.06403}{{\ttfamily
  1804.06403}}].

\bibitem{Mishra:2019}
N.~Mishra and E.~Schaan, \emph{{Bias to CMB lensing from lensed foregrounds}},
  \href{https://doi.org/10.1103/PhysRevD.100.123504}{\emph{\prd} {\bfseries
  100} (2019) 123504} [\href{https://arxiv.org/abs/1908.08057}{{\ttfamily
  1908.08057}}].

\bibitem{Sailer:2020}
N.~Sailer, E.~Schaan and S.~Ferraro, \emph{{Lower bias, lower noise CMB lensing
  with foreground-hardened estimators}},
  \href{https://doi.org/10.1103/PhysRevD.102.063517}{\emph{\prd} {\bfseries
  102} (2020) 063517} [\href{https://arxiv.org/abs/2007.04325}{{\ttfamily
  2007.04325}}].

\bibitem{Han:2021:FG}
D.~Han and N.~Sehgal, \emph{{Mitigating foreground bias to the CMB lensing
  power spectrum for a CMB-HD survey}},
  \href{https://doi.org/10.1103/PhysRevD.105.083516}{\emph{\prd} {\bfseries
  105} (2022) 083516} [\href{https://arxiv.org/abs/2112.02109}{{\ttfamily
  2112.02109}}].

\bibitem{Darwish:2021}
O.~Darwish, B.D.~Sherwin, N.~Sailer, E.~Schaan and S.~Ferraro,
  \emph{{Optimizing foreground mitigation for CMB lensing with combined
  multifrequency and geometric methods}},
  \href{https://doi.org/10.1103/PhysRevD.107.043519}{\emph{\prd} {\bfseries
  107} (2023) 043519}.

\bibitem{Sailer:2022}
N.~Sailer, S.~Ferraro and E.~Schaan, \emph{{Foreground-immune CMB lensing
  reconstruction with polarization}},
  \href{https://doi.org/10.1103/PhysRevD.107.023504}{\emph{{\prd}} (2022) }
  [\href{https://arxiv.org/abs/2211.03786}{{\ttfamily 2211.03786}}].

\bibitem{Qu:2022:shear}
F.J.~Qu, A.~Challinor and B.D.~Sherwin, \emph{{CMB lensing with shear-only
  reconstruction on the full sky}},
  \href{https://doi.org/10.1103/PhysRevD.108.063518}{\emph{\prd} {\bfseries
  108} (2023) 063518} [\href{https://arxiv.org/abs/2208.14988}{{\ttfamily
  2208.14988}}].

\bibitem{Mirmelstein:2020:sys}
M.~Mirmelstein, G.~Fabbian, A.~Lewis and J.~Peloton, \emph{{Instrumental
  systematics biases in CMB lensing reconstruction: A simulation-based
  assessment}}, \href{https://doi.org/10.1103/PhysRevD.103.123540}{\emph{\prd}
  {\bfseries 103} (2021) 123540}
  [\href{https://arxiv.org/abs/2011.13910}{{\ttfamily 2011.13910}}].

\bibitem{Nagata:2021:sys}
R.~Nagata and T.~Namikawa, \emph{{A numerical study of observational systematic
  errors in lensing analysis of CMB polarization}},
  \href{https://doi.org/10.1093/ptep/ptab040}{\emph{\ptep} {\bfseries 2021}
  (2021) 053} [\href{https://arxiv.org/abs/2102.00133}{{\ttfamily
  2102.00133}}].

\bibitem{Beck:2020:lens-FG}
D.~Beck, J.~Errard and R.~Stompor, \emph{{Impact of Polarized Galactic
  Foreground Emission on CMB Lensing Reconstruction and Delensing of B-Modes}},
  \href{https://doi.org/10.1088/1475-7516/2020/06/030}{\emph{\jcap} {\bfseries
  06} (2020) 030} [\href{https://arxiv.org/abs/2001.02641}{{\ttfamily
  2001.02641}}].

\bibitem{LiteBIRD:lens:paper2}
T.~Namikawa, A.~Lonappan et~al., \emph{{}}, {\emph{in prep.} (2023) }.

\bibitem{P18:cosmological-parameters}
{\textit{Planck} Collaboration}, \emph{{Planck 2018 results. VI. Cosmological
  parameters}}, \href{https://doi.org/10.1051/0004-6361/201833910}{\emph{\aap}
  {\bfseries 641} (2020) A6}
  [\href{https://arxiv.org/abs/1807.06209}{{\ttfamily 1807.06209}}].

\bibitem{Tegmark:2003}
M.~Tegmark, A.~de~Oliveira-Costa and A.J.S.~Hamilton, \emph{{High resolution
  foreground cleaned {CMB} map from {WMAP}}},
  \href{https://doi.org/10.1103/physrevd.68.123523}{\emph{\prd} {\bfseries 68}
  (2003) }.

\bibitem{Challinor:2002cd}
A.~Challinor and G.~Chon, \emph{{Geometry of weak lensing of CMB
  polarization}}, \href{https://doi.org/10.1103/PhysRevD.66.127301}{\emph{Phys.
  Rev. D} {\bfseries 66} (2002) 127301}
  [\href{https://arxiv.org/abs/astro-ph/0301064}{{\ttfamily
  astro-ph/0301064}}].

\bibitem{Lewis:2006:review}
A.~Lewis and A.~Challinor, \emph{{Weak gravitational lensing of the CMB}},
  \href{https://doi.org/10.1016/j.physrep.2006.03.002}{\emph{Phys. Rep.}
  {\bfseries 429} (2006) 1}
  [\href{https://arxiv.org/abs/astro-ph/0601594}{{\ttfamily
  astro-ph/0601594}}].

\bibitem{Hanson:2009:review}
D.~Hanson, A.~Challinor and A.~Lewis, \emph{{Weak lensing of the CMB}},
  \href{https://doi.org/10.1007/s10714-010-1036-y}{\emph{Gen. Rel. Grav.}
  {\bfseries 42} (2010) 2197}
  [\href{https://arxiv.org/abs/0911.0612}{{\ttfamily 0911.0612}}].

\bibitem{Hadzhiyska:2017nqe}
B.~Hadzhiyska, D.~Spergel and J.~Dunkley, \emph{{Small-scale modification to
  the lensing kernel}},
  \href{https://doi.org/10.1103/PhysRevD.97.043521}{\emph{\prd} {\bfseries 97}
  (2018) 043521} [\href{https://arxiv.org/abs/1711.03168}{{\ttfamily
  1711.03168}}].

\bibitem{Pratten:2016}
G.~Pratten and A.~Lewis, \emph{Impact of post-born lensing on the cmb},
  \href{https://doi.org/10.48550/arXiv.1605.05662}{\emph{\jcap} {\bfseries 08}
  (2016) 047} [\href{https://arxiv.org/abs/1605.05662}{{\ttfamily
  1605.05662}}].

\bibitem{Lewis:Pratten:2016}
A.~Lewis and G.~Pratten, \emph{{Effect of lensing non-Gaussianity on the CMB
  power spectra}},
  \href{https://doi.org/10.1088/1475-7516/2016/12/003}{\emph{\jcap} {\bfseries
  12} (2016) 003} [\href{https://arxiv.org/abs/1608.01263}{{\ttfamily
  1608.01263}}].

\bibitem{Fabbian:2017wfp}
G.~Fabbian, M.~Calabrese and C.~Carbone, \emph{{CMB weak-lensing beyond the
  Born approximation: a numerical approach}},
  \href{https://doi.org/10.1088/1475-7516/2018/02/050}{\emph{\jcap} {\bfseries
  1802} (2018) 050} [\href{https://arxiv.org/abs/1702.03317}{{\ttfamily
  1702.03317}}].

\bibitem{Namikawa:2014:review}
T.~Namikawa, \emph{{Cosmology from weak lensing of CMB}},
  \href{https://doi.org/10.1093/ptep/ptu044}{\emph{\ptep} {\bfseries 2014}
  (2014) 06B108} [\href{https://arxiv.org/abs/1403.3569}{{\ttfamily
  1403.3569}}].

\bibitem{Zaldarriaga:1998ar}
M.~Zaldarriaga and U.~Seljak, \emph{{Gravitational lensing effect on cosmic
  microwave background polarization}},
  \href{https://doi.org/10.1103/PhysRevD.58.023003}{\emph{Phys. Rev. D}
  {\bfseries 58} (1998) 023003}
  [\href{https://arxiv.org/abs/astro-ph/9803150}{{\ttfamily
  astro-ph/9803150}}].

\bibitem{Lewis:2011:bispec}
A.~Lewis, A.~Challinor and D.~Hanson, \emph{{The shape of the CMB lensing
  bispectrum}},
  \href{https://doi.org/10.1088/1475-7516/2011/03/018}{\emph{\jcap} {\bfseries
  03} (2011) 018} [\href{https://arxiv.org/abs/1101.2234}{{\ttfamily
  1101.2234}}].

\bibitem{Komatsu:2022:review}
E.~Komatsu, \emph{{New physics from the polarized light of the cosmic microwave
  background}}, \href{https://doi.org/10.1038/s42254-022-00452-4}{\emph{Nature
  Rev. Phys.} {\bfseries 4} (2022) 452}
  [\href{https://arxiv.org/abs/2202.13919}{{\ttfamily 2202.13919}}].

\bibitem{Namikawa:2021gbr}
T.~Namikawa, \emph{{CMB mode coupling with isotropic polarization rotation}},
  \href{https://doi.org/10.1093/mnras/stab1796}{\emph{Mon. Not. Roy. Astron.
  Soc.} {\bfseries 506} (2021) 1250}
  [\href{https://arxiv.org/abs/2105.03367}{{\ttfamily 2105.03367}}].

\bibitem{Smith_PS}
K.M.~Smith et~al., \emph{{CMBPol Mission Concept Study: Gravitational
  Lensing}}, \href{https://doi.org/10.1063/1.3160886}{\emph{AIP Conf. Proc.}
  {\bfseries 1141} (2009) 121}
  [\href{https://arxiv.org/abs/0811.3916}{{\ttfamily 0811.3916}}].

\bibitem{macellari_FF}
N.~Macellari, E.~Pierpaoli, C.~Dickinson and J.E.~Vaillancourt, \emph{{Galactic
  foreground contributions to the 5-year Wilkinson Microwave Anisotropy Probe
  maps}}, \href{https://doi.org/10.1111/j.1365-2966.2011.19542.x}{\emph{Monthly
  Notices of the Royal Astronomical Society} {\bfseries 418} (2011) 888}
  [\href{https://arxiv.org/abs/https://academic.oup.com/mnras/article-pdf/418/2/888/3691294/mnras0418-0888.pdf}{{\ttfamily
  https://academic.oup.com/mnras/article-pdf/418/2/888/3691294/mnras0418-0888.pdf}}].

\bibitem{Delabrouille:2008:NILC}
J.~Delabrouille, J.F.~Cardoso, M.L.~Jeune, M.~Betoule, G.~Fay and F.~Guilloux,
  \emph{{A full sky, low foreground, high resolution CMB map from WMAP}},
  \href{https://doi.org/10.1051/0004-6361:200810514}{\emph{\aap} {\bfseries
  493} (2009) 835} [\href{https://arxiv.org/abs/0807.0773}{{\ttfamily
  0807.0773}}].

\bibitem{Eriksen:2004:wiener}
H.K.~Eriksen, I.J.~O'Dwyer, J.B.~Jewell, B.D.~Wandelt, D.L.~Larson, K.M.~Gorski
  et~al., \emph{{Power spectrum estimation from high-resolution maps by Gibbs
  sampling}}, \href{https://doi.org/10.1086/425219}{\emph{Astrophys. J. Suppl.}
  {\bfseries 155} (2004) 227}
  [\href{https://arxiv.org/abs/astro-ph/0407028}{{\ttfamily
  astro-ph/0407028}}].

\bibitem{Namikawa:2021:SO-delens}
T.~Namikawa, A.~{Baleato Lizancos}, N.~Robertson, B.D.~Sherwin, A.~Challinor,
  D.~Alonso et~al., \emph{{Simons Observatory: Constraining inflationary
  gravitational waves with multitracer B-mode delensing}},
  \href{https://doi.org/10.1103/PhysRevD.105.023511}{\emph{\prd} {\bfseries
  105} (2022) 023511} [\href{https://arxiv.org/abs/2110.09730}{{\ttfamily
  2110.09730}}].

\bibitem{Remazeilles:2014:s1model}
M.~Remazeilles, C.~Dickinson, A.J.~Banday, M.A.~Bigot-Sazy and T.~Ghosh,
  \emph{{An improved source-subtracted and destriped 408 MHz all-sky map}},
  \href{https://doi.org/10.1093/mnras/stv1274}{\emph{\mnras} {\bfseries 451}
  (2015) 4311} [\href{https://arxiv.org/abs/1411.3628}{{\ttfamily 1411.3628}}].

\bibitem{Bennett:2013:wmap}
C.L.~{Bennett}, D.~{Larson}, J.L.~{Weiland}, N.~{Jarosik}, G.~{Hinshaw},
  N.~{Odegard} et~al., \emph{{Nine-year Wilkinson Microwave Anisotropy Probe
  (WMAP) Observations: Final Maps and Results}},
  \href{https://doi.org/10.1088/0067-0049/208/2/20}{\emph{\apjs} {\bfseries
  208} (2013) 20} [\href{https://arxiv.org/abs/1212.5225}{{\ttfamily
  1212.5225}}].

\bibitem{Miville-Deschenes:2008:FG}
M.A.~Miville-Deschenes, N.~Ysard, A.~Lavabre, N.~Ponthieu, J.F.~Macias-Perez,
  J.~Aumont et~al., \emph{{Separation of anomalous and synchrotron emissions
  using WMAP polarization data}},
  \href{https://doi.org/10.1051/0004-6361:200809484}{\emph{\aap} {\bfseries
  490} (2008) 1093} [\href{https://arxiv.org/abs/0802.3345}{{\ttfamily
  0802.3345}}].

\bibitem{P15:FG}
{\textit{Planck} Collaboration}, \emph{{Planck 2015 results. X. Diffuse
  component separation: Foreground maps}},
  \href{https://doi.org/10.1051/0004-6361/201525967}{\emph{\aap} {\bfseries
  594} (2016) A10} [\href{https://arxiv.org/abs/1502.01588}{{\ttfamily
  1502.01588}}].

\bibitem{gorski}
K.M.~{G{\'o}rski}, E.~{Hivon}, A.J.~{Banday}, B.D.~{Wandelt}, F.K.~{Hansen},
  M.~{Reinecke} et~al., \emph{{HEALPix: A Framework for High-Resolution
  Discretization and Fast Analysis of Data Distributed on the Sphere}},
  \href{https://doi.org/10.1086/427976}{\emph{\apj} {\bfseries 622} (2005) 759}
  [\href{https://arxiv.org/abs/arXiv:astro-ph/0409513}{{\ttfamily
  arXiv:astro-ph/0409513}}].

\bibitem{Kesden:2003:N1}
M.H.~Kesden, A.~Cooray and M.~Kamionkowski, \emph{{Lensing reconstruction with
  CMB temperature and polarization}},
  \href{https://doi.org/10.1103/PhysRevD.67.123507}{\emph{\prd} {\bfseries 67}
  (2003) 123507} [\href{https://arxiv.org/abs/astro-ph/0302536}{{\ttfamily
  astro-ph/0302536}}].

\bibitem{cmblensplus}
T.~Namikawa, ``{cmblensplus: A tool to analyze cosmic microwave background
  anisotropies}.'' Astrophysics Source Code Library, record ascl:2104.021,
  Apr., 2021.

\bibitem{Hanson:2010:N2}
D.~Hanson, A.~Challinor, G.~Efstathiou and P.~Bielewicz, \emph{Cmb temperature
  lensing power reconstruction},
  \href{https://doi.org/10.1103/PhysRevD.83.043005}{\emph{\prd} {\bfseries 83}
  (2011) 043005} [\href{https://arxiv.org/abs/1008.4403}{{\ttfamily
  1008.4403}}].

\bibitem{Boehm:2016}
V.~Boehm, M.~Schmittfull and B.~Sherwin, \emph{{Bias to CMB lensing
  measurements from the bispectrum of large-scale structure}},
  \href{https://doi.org/10.1103/PhysRevD.94.043519}{\emph{\prd} {\bfseries 94}
  (2016) 043519} [\href{https://arxiv.org/abs/1605.01392}{{\ttfamily
  1605.01392}}].

\bibitem{Boehm:2018}
V.~Boehm, B.D.~Sherwin, J.~Liu, J.C.~Hill, M.~Schmittfull and T.~Namikawa,
  \emph{On the effect of non-gaussian lensing deflections on cmb lensing
  measurements}, \href{https://doi.org/10.1103/PhysRevD.98.123510}{\emph{\prd}
  {\bfseries 98} (2018) 123510}
  [\href{https://arxiv.org/abs/1806.01157}{{\ttfamily 1806.01157}}].

\bibitem{Schmittfull:2013}
M.M.~Schmittfull, A.~Challinor, D.~Hanson and A.~Lewis, \emph{{Joint analysis
  of CMB temperature and lensing-reconstruction power spectra}},
  \href{https://doi.org/10.1103/PhysRevD.88.063012}{\emph{\prd} {\bfseries 88}
  (2013) 063012} [\href{https://arxiv.org/abs/1308.0286}{{\ttfamily
  1308.0286}}].

\bibitem{Peloton:2016:RDN0}
J.~Peloton, M.~Schmittfull, A.~Lewis, J.~Carron and O.~Zahn, \emph{{Full
  covariance of CMB and lensing reconstruction power spectra}},
  \href{https://doi.org/10.1103/PhysRevD.95.043508}{\emph{\prd} {\bfseries 95}
  (2017) 043508} [\href{https://arxiv.org/abs/1611.01446}{{\ttfamily
  1611.01446}}].

\bibitem{Fabbian:2019:cross}
G.~Fabbian, A.~Lewis and D.~Beck, \emph{{CMB lensing reconstruction biases in
  cross-correlation with large-scale structure probes}},
  \href{https://doi.org/10.1088/1475-7516/2019/10/057}{\emph{\jcap} {\bfseries
  10} (2019) 057} [\href{https://arxiv.org/abs/1906.08760}{{\ttfamily
  1906.08760}}].

\bibitem{EuclidForecastIST:2020}
{Euclid Collaboration}, A.~{Blanchard}, S.~{Camera}, C.~{Carbone},
  V.F.~{Cardone}, S.~{Casas} et~al., \emph{{Euclid preparation. VII. Forecast
  validation for Euclid cosmological probes}},
  \href{https://doi.org/10.1051/0004-6361/202038071}{\emph{\aap} {\bfseries
  642} (2020) A191} [\href{https://arxiv.org/abs/1910.09273}{{\ttfamily
  1910.09273}}].

\bibitem{LSSTCollaboration:2018}
{The LSST Dark Energy Science Collaboration}, R.~Mandelbaum, T.~Eifler,
  R.~Hložek, T.~Collett, E.~Gawiser et~al., \emph{The lsst dark energy science
  collaboration (desc) science requirements document},
  \href{https://doi.org/10.48550/ARXIV.1809.01669}{\emph{{}} (2018) }.

\bibitem{Dalal:2007:fNL}
N.~Dalal, O.~Dore, D.~Huterer and A.~Shirokov, \emph{{The imprints of
  primordial non-gaussianities on large-scale structure: scale dependent bias
  and abundance of virialized objects}},
  \href{https://doi.org/10.1103/PhysRevD.77.123514}{\emph{\prd} {\bfseries 77}
  (2008) 123514} [\href{https://arxiv.org/abs/0710.4560}{{\ttfamily
  0710.4560}}].

\bibitem{Jeong:2009:fNL}
D.~Jeong, E.~Komatsu and B.~Jain, \emph{{Galaxy-CMB and galaxy-galaxy lensing
  on large scales: sensitivity to primordial non-Gaussianity}}, {\emph{\prd}
  {\bfseries 80} (2009) 123527}
  [\href{https://arxiv.org/abs/0910.1361}{{\ttfamily 0910.1361}}].

\bibitem{Schmittfull:2017:fNL}
M.~Schmittfull and U.~Seljak, \emph{{Parameter constraints from
  cross-correlation of CMB lensing with galaxy clustering}},
  \href{https://doi.org/10.1103/PhysRevD.97.123540}{\emph{\prd} {\bfseries 97}
  (2018) 123540} [\href{https://arxiv.org/abs/1710.09465}{{\ttfamily
  1710.09465}}].

\bibitem{Ballardini:2019:fNL}
M.~Ballardini, W.L.~Matthewson and R.~Maartens, \emph{{Constraining primordial
  non-Gaussianity using two galaxy surveys and CMB lensing}},
  \href{https://doi.org/10.1093/mnras/stz2258}{\emph{\mnras} {\bfseries 489}
  (2019) 1950} [\href{https://arxiv.org/abs/1906.04730}{{\ttfamily
  1906.04730}}].

\bibitem{McCarthy:2020:cib}
F.~McCarthy and M.S.~Madhavacheril, \emph{{Improving models of the cosmic
  infrared background using CMB lensing mass maps}},
  \href{https://doi.org/10.1103/PhysRevD.103.103515}{\emph{\prd} {\bfseries
  103} (2021) 103515} [\href{https://arxiv.org/abs/2010.16405}{{\ttfamily
  2010.16405}}].

\bibitem{McCarthy:2022:fNL}
F.~McCarthy, M.S.~Madhavacheril and A.S.~Maniyar, \emph{{Constraints on
  primordial non-Gaussianity from halo bias measured through CMB lensing
  cross-correlations}},
  \href{https://doi.org/10.1103/PhysRevD.108.083522}{\emph{\prd} {\bfseries
  108} (2023) 083522} [\href{https://arxiv.org/abs/2210.01049}{{\ttfamily
  2210.01049}}].

\bibitem{Sachs:1967}
R.K.~Sachs and A.M.~Wolfe, \emph{{Perturbations of a Cosmological Model and
  Angular Variations of the Microwave Background}},
  \href{https://doi.org/10.1086/148982}{\emph{\apj} {\bfseries 147} (1967) 73}.

\bibitem{Goldberg:1999xm}
D.M.~Goldberg and D.N.~Spergel, \emph{{Microwave background bispectrum. 2. A
  probe of the low redshift universe}}, {\emph{\prd} {\bfseries 59} (1999)
  103002} [\href{https://arxiv.org/abs/astro-ph/9811251}{{\ttfamily
  astro-ph/9811251}}].

\bibitem{Pearson:2014}
R.~Pearson, B.~Sherwin and A.~Lewis, \emph{{CMB lensing reconstruction using
  cut sky polarization maps and pure-$B$ modes}},
  \href{https://doi.org/10.1103/PhysRevD.90.023539}{\emph{Phys. Rev. D}
  {\bfseries 90} (2014) 023539}
  [\href{https://arxiv.org/abs/1403.3911}{{\ttfamily 1403.3911}}].

\bibitem{Mirmelstein:2019:filter}
M.~Mirmelstein, J.~Carron and A.~Lewis, \emph{{Optimal filtering for CMB
  lensing reconstruction}},
  \href{https://doi.org/10.1103/PhysRevD.100.123509}{\emph{\prd} {\bfseries
  100} (2019) 123509} [\href{https://arxiv.org/abs/1909.02653}{{\ttfamily
  1909.02653}}].

\end{thebibliography}\endgroup

\end{document}